\documentclass[10pt,journal]{IEEEtran}
\usepackage{amsthm}
\usepackage{amsmath}
\usepackage[normalem]{ulem}
\usepackage{soul} 
\usepackage{paralist}
\usepackage{xcolor}
\usepackage{balance}

\usepackage{amssymb}
\usepackage{graphicx}
\usepackage{epstopdf}
\usepackage{subfig}
\usepackage{multirow}
\usepackage{enumerate}
\usepackage[hidelinks]{hyperref}
\hypersetup{pdftitle=timing,pdfdisplaydoctitle}

\theoremstyle{definition}

\def\etc{\emph{etc}}

\def\etc{\emph{etc}}
\def\eg{\emph{e.g. }}

\newcommand{\vv}[1]{\boldsymbol{#1}}
\newcommand{\specialcell}[2][c]{%
  \begin{tabular}[#1]{@{}c@{}}#2\end{tabular}}

\begin{document}

\title{A Web Traffic Analysis Attack Using Only Timing Information}
\author{\IEEEauthorblockN{Saman Feghhi,
Douglas J. Leith}\thanks{Copyright (c) 2016 IEEE. Personal use of this material is permitted. However, permission to use this material for any other purposes must be obtained from the IEEE by sending a request to pubs-permissions@ieee.org}\thanks{This work was supported by Science Foundation Ireland under Grant No. 11/PI/1177.}\\
\IEEEauthorblockA{School of Computer Science and Statistics\\Trinity College Dublin}\\
Email: \{feghhis,doug.leith\}@tcd.ie}

\maketitle

\begin{abstract}
We introduce an attack against encrypted web traffic that makes use only of packet timing information on the uplink.   This attack is therefore impervious to existing packet padding defences.  In addition, unlike existing approaches this timing-only attack does not require knowledge of the start/end of web fetches and so is effective against traffic streams.   We demonstrate the effectiveness of the attack against both wired and wireless traffic, achieving mean success rates in excess of 90\%.   In addition to being of interest in its own right, this timing-only attack serves to highlight deficiencies in existing defences and so to areas where it would be beneficial for Virtual Private Network (VPN) designers to focus further attention.
\end{abstract}

\begin{IEEEkeywords}
traffic analysis, website fingerprinting, timing-only attacks, network privacy.
\end{IEEEkeywords}

\section{Introduction}
\begin{figure}
\centering
\includegraphics[width=0.8\columnwidth]{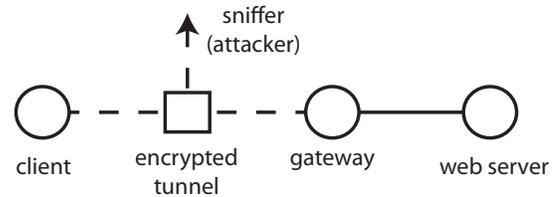}
\caption{Schematic illustrating attacker of the type considered.  A client machine is connected to an external network via an encrypted tunnel (ssh, SSL, IPSec \etc.).  The attacker can detect the time when packets traverse the tunnel in the uplink direction, but has no other information about the clients activity.}\label{fig:intoexample}
\end{figure}

\noindent
In this paper we consider an attacker of the type illustrated in Figure \ref{fig:intoexample}.  The attacker can detect the time when packets traverse the encrypted tunnel in the uplink direction, but has no other information about the clients' activity.   The attacker's objective is to use this information to guess, with high probability of success, the web sites which the client visits. What is distinctive about the attack considered here is that the attacker relies solely on packet timestamp information whereas the previously reported attacks against encrypted web traffic have mainly made use of observations of packet size and/or packet count information.    

Our interest in timing-only attacks is twofold.   Firstly, packet padding is a relatively straightforward defence against attacks that rely primarily on packet size, and indeed is currently either already available or being implemented in a number of popular VPNs \cite{cai12}.  Secondly, alternative attacks based on packet counting \cite{cai12,dyer12} are insensitive to packet padding defences but require partitioning of a packet stream into individual web fetches in order for the number of packets associated with each web fetch to be determined, which may be highly challenging in practice on links where there are no clear pauses between web fetches.   In contrast, packet timing-based attacks are not only largely unaffected by packet padding defences but also, as we will show, do not require partitioning of the packet stream.   Hence, they are potentially a practically important class of attack against current and future VPNs.   While some work has been carried out using inter-arrival time information to classify the application (HTTP, IMAP \etc.) \cite{jaber11}, to our knowledge, there is no previous work reporting use of timing information alone to construct a successful attack against encrypted web traffic.

The main contributions of the present paper are as follows: (i) we describe an attack against encrypted web traffic that uses packet timing information alone, (ii) we demonstrate that this attack is highly effective against both wired and wireless traffic, achieving mean success rates in excess of 90\% over ethernet and wireless tunnels and a success rate of 58\% against Tor traffic, (iii) we also demonstrate that the attack is effective against traffic streams \emph{i.e.} back to back web page fetches where the packet boundaries between fetches are unknown.

In addition to being of interest in its own right, particularly in view of the powerful nature of the attack, this timing-only attack also serves to highlight deficiencies in existing defences and so to areas where it would be beneficial for VPN designers to focus further attention.    We note that, complementary to the present work, in \cite{dyer12} it is demonstrated that when the web fetch boundaries within a packet stream are known then an NGRAM approach using packet count together with uplink/downlink direction information is also sufficient to construct an effective attack against encrypted web traffic despite packet padding.   Hence, we can conclude that (i) uplink/downlink packet ordering plus web fetch boundaries and (ii) uplink/downlink packet timing information are both sensitive quantities that ought to be protected by a secure encrypted tunnel.  Packet padding does not protect these quantities.   Directing defences against these two sets of packet stream features therefore seems an important direction for future work.

\section{Related Work}
\noindent
The general topic of traffic analysis has been the subject of much interest, and a large body of literature exists.  Some of the earliest work specifically focussed on attacks and defences for encrypted web traffic appears to be that of Hintz \cite{hintz02}, which considers the SafeWeb encrypting proxy.   In this setup (i)  web page fetches occur sequentially with the start and end of each web page fetch known, and for each packet (ii) the client-side port number, (iii) the direction (incoming/outgoing) and (iv) the size is observed.   A web page signature is constructed consisting of the aggregate bytes received on each port (calculated by summing packet sizes), effectively corresponding to the number and size of each object within the web page.   In \cite{sun03} it is similarly assumed that the number and size of the objects in a web page can be observed and using this information a classification success rate of 75\% is reported. 

Subsequently,  Bissias \emph{et al} \cite{bissias06} considered an encrypted tunnel setup where (i) web page fetches occur sequentially with the start and end of each web page fetch known, and for each packet (ii) the size, (iii) the direction (incoming/outgoing) and (iv) the time (and so also the packet ordering)  is observed.    The sequence of packet inter-arrival times and packet sizes from a web page fetch is used to create a profile for each web page in a target set and the cross correlation between an observed traffic sequence and the stored profiles is then used as a measure of similarity.   A classification accuracy of 23\% is observed when using a set of 100 web pages, rising to 40\% when restricted to a smaller set of web pages. 

Most later work has adopted essentially the same model as \cite{bissias06}, making use of packet direction and size information and assuming that the packet stream has already been partitioned into individual web page fetches.  For example in \cite{tao14} the timing information is not considered in the feature set, hence the attack can be countered with defences such as BuFLO in \cite{dyer12} leading to a success rate of only $10\%$. In  \cite{liberatore06,herrmann09} Bayes  classifiers based on the direction and size of packets are considered while in \cite{panchenko11} an SVM classifier is proposed.  In \cite{lu10} classification based on direction and size of packets is studied using Levenshtein distance as the similarity metric, in \cite{miller14} using a Gaussian Bag-of-Words approach and in \cite{tao14} using $K$-NN classification.  In \cite{cai12} using a SVM approach a classification accuracy of over 80\% is reported for both SSH and Tor traffic and the  defences considered were generally found to be ineffective.   Similarly, \cite{dyer12} considers Bayes and SVM classifiers and finds that a range of proposed defences are ineffective.  In \cite{gong10} remote inference of packet sizes from queueing delay is studied.

\section{Anatomy of a Web Page Fetch}
\noindent
When traffic is carried over an encrypted tunnel, such as a VPN, the packet source and destination addresses and ports and the packet payload are hidden.  We also assume here that the tunnel pads the packets to be of equal size, so that packet size information is also concealed, and that the start and end of an individual web fetch may also be concealed \emph{e.g.} when the web fetch is embedded in a larger traffic stream.   An attacker sniffing traffic on the encrypted tunnel is therefore able only to observe the direction and timing of packets through the tunnel, \emph{i.e.} to observe a sequence of pairs $\{(t_k,d_k)\}$, $k=1,2,\cdots$ where $t_k$ is the time at which the $k$-th packet is observed and $d_k\in\{-1,1\}$ indicates whether the packet is travelling in the uplink or downlink direction.   Our experiments on use of uplink, downlink and uplink+downlink traffic suggest that downlink traffic provides no additional information regarding timing patterns over uplink traffic. The reason is that the timing of ACKs in uplink traffic is correlated to that of downlink packets which means that using only uplink traffic provides sufficient information. Furthermore it may be easier for an eavesdropper to access unmodified uplink traffic on the first hop, (given the traffic comes immediately from the source, while the corresponding downlink traffic could be morphed using inter-flow transformations \eg flow mixing, split and merge \cite{wang07}). We therefore focus on an attacker that can only observe the timestamps $\{t_k\}$, $k\in K_{up}:=\{\kappa\in\{1,2,\cdots\}:d_\kappa=-1\}$ associated with uplink traffic.

\begin{figure}
\centering
	\includegraphics[width=0.8\columnwidth]{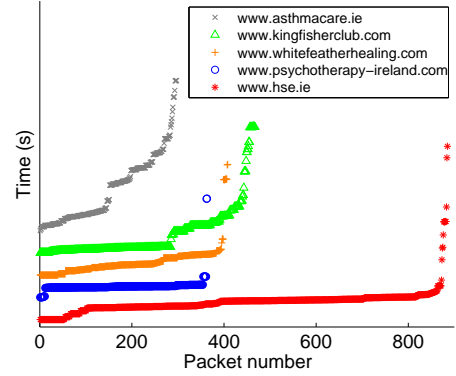}
	\caption{Time traces of uplink traffic from 5 different Irish health-related web sites are shown. It can be seen that the web site time traces exhibit distinct patterns.  The traces are shifted vertically to avoid overlap and facilitate comparison.}
	\label{fig:difweb}
\end{figure}

Figure \ref{fig:difweb} plots the timestamps $\{t_k\}$ of the uplink packets sent during the course of fetching five different health-related web pages (see below for details of the measurement setup).    The $x$-axis indicates the packet number $k$ within the stream and the $y$-axis the corresponding timestamp $t_k$ in seconds.   It can be seen that these timestamp traces are distinctly different for each web site, and it is this observation that motivates interest in whether timing analysis may by itself (without additional information such as packet size, uplink/downlink packet ordering \etc.) be sufficient to successfully de-anomymise encrypted web traffic.

\begin{figure*}
\centering
	\includegraphics[width=1.0\textwidth]{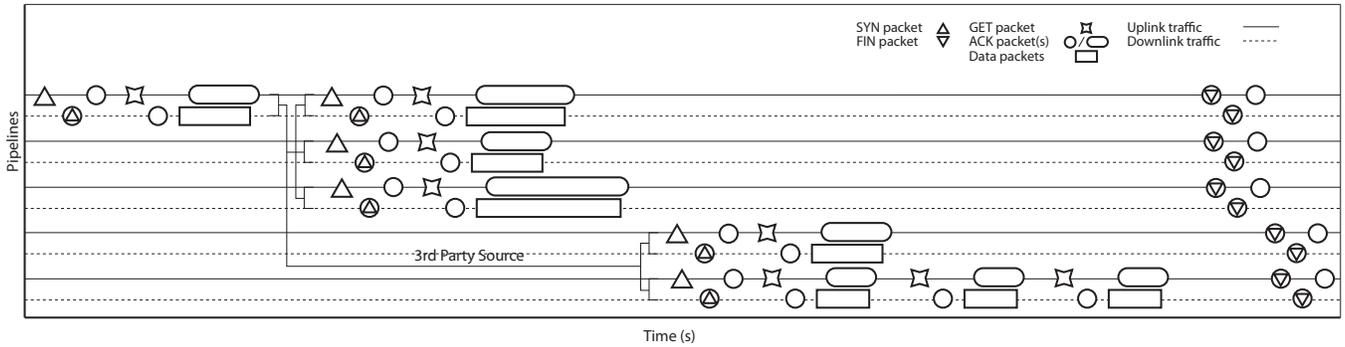}
	\caption{This figure represent a typical web site query. It starts by requesting the index page. Then as the browser parses through this page more objects are fetched in parallel. Some objects may also be outsourced to 3rd party web sites which have their own pipelines. Dynamic content may be updated at intervals, as indicated in the last two lines of the figure, and connections tend to close in groups.}
	\label{fig:pstream}
\end{figure*}

To gain insight into the differences between the packet timestamp sequences in Figure \ref{fig:difweb}  and, importantly, whether they are genuinely related to characteristics of each web page rather than to other factors, it is helpful to consider the process of fetching a web page in more detail.    To fetch a web page the client browser starts by opening a TCP connection with the server indicated by the URL and issues an HTTP GET or POST request to which the server then replies.  As the client parses the server response it issues additional GET/POST requests to fetch embedded objects (images, css, scripts \etc.).   These additional requests may be to different servers from the original request (\emph{e.g.} when the object to be fetched is an advert or is hosted in a separate content-delivery network),  in which case the client opens a TCP connection to each new server in order to issue the requests.   Fetching of these objects may in turn trigger the fetching of further objects.   Note that asynchronous fetching of dynamic content using, \emph{e.g.} AJAX, can lead to a complex sequence of server requests and responses even after the page has been rendered by the browser.   Also, typically the TCP connections to the various servers are held open until the page is fully loaded so that they can be reused for later requests (request pipelining in this way is almost universally used by modern browsers).  

This web fetch process is illustrated schematically in Figure \ref{fig:pstream}.   We make the following  more detailed observations:
\begin{enumerate}
\item \emph{Connection to third-party servers}.   Fetching an object located on a third-party server requires the opening of a new TCP connection to that server, over which the HTTP request is then sent.   The TCP connection handshake introduces a delay (of at least one RTT) and since the pattern of these delays is related to the web page content it can potentially assist in identifying the web page.  

\item \emph{Pipelining of requests}.   Multiple objects located on the same server lead to several GET/POST requests being sent to that server, one after another.   Due to the dynamics of TCP congestion control, this burst of back-to-back requests can affect the timing of the response packets in a predictable manner that once again can potentially assist in identifying the web page.   

\item \emph{Asynchronous requests}.  Dynamic content, \emph{e.g.} pre-fetching via AJAX, can lead to update requests to a server with large inter-arrival times that can potentially act as a web page signature. 

\item \emph{Connection closing.}  When a web page fetch is completed, the associated TCP connections are closed.   A FIN/FINACK/ACK exchange closes each connection and this burst of packets can have quite distinctive timing which allows it to be identified.    Since the number of connections is related to the number of distinct locations where objects in the web page are stored, it changes between web pages.   
\end{enumerate}
Our aim is to use timing features such as these, which vary depending upon the web page fetched, to create a timing signature which allows us to identify which web page is being fetched based on timing data only.

\section{Comparing Sequences of Packet Timestamps}\label{sec:distance}
\noindent
Suppose we have two sequences of packet timestamps $\vv{t}:=\{t_i\}$, $ i=1,2,\cdots,n$ and $\vv{t}^\prime:=\{t_j^\prime\}$, $j=1,2,\cdots,m$.  Note that for simplicity we re-label the uplink packet indices to start from $1$ and to increase consecutively since none of our analysis will depend on this.   Note also that the sequence lengths $n$ and $m$ are \emph{not} assumed to be the same.   To proceed we need to define an appropriate measure of the distance between such sequences.

\subsection{Network Distortion of Timestamp Sequences}
\noindent
The packet stream observed during a web page fetch is affected by network events during the fetch.   Changes in download rate (\emph{e.g.} due to flows starting/finishing within the network) tend to  stretch/compress the times between packets.   Queueing within the network also affects packet timing, with queued packets experiencing both greater delay and tending to be more bunched together.  Link-layer retransmission on wireless links has a similar effect to queueing.   Similarly to changes in download rate, the effect is primarily to stretch/compress the times between packets.   

\begin{figure}
\centering
\includegraphics[width=0.8\columnwidth]{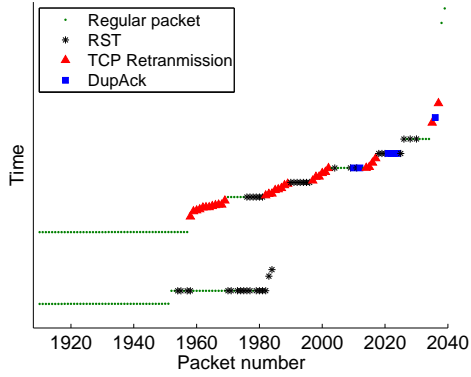}
\caption{Illustrating impact of changes in packet loss on the packet timestamp  sequence. The bottom sequence shows the packet sequence at connection closing of a loss-free web fetch, while the top sequence shows the corresponding section from a different fetch of the same web page that was subject to packet loss and exhibits TCP retransmissions and DupACKs.}\label{fig:bw}
\end{figure}

Packet loss introduces a ``hole'' in the packet stream where the packet ought to have arrived and also affects the timing of later packets due to the action of TCP congestion control (which reduces the send rate on packet loss) and retransmission of the lost packets.  For example, Figure \ref{fig:bw} shows uplink measurements of packet retransmissions and duplicate ACKs at the end of two fetches of the same web page where it can be seen that these have the effect of stretching the packet sequence.

\subsection{Derivative Dynamic Time Warping}
\noindent
Our interest is in a measure of the distance between packet sequences which is insensitive to the types of distortion introduced by the network, so that the distance between packet streams $\vv{t}$ and $\vv{t}^\prime$ associated with fetches of the same web page at different times is measured as being small, and ideally the distance between fetches of different web pages is measured to be large.  To this end we use a variant of Dynamic Time Warping (DTW) \cite{keogh01}.    DTW aims to be insensitive to differences between sequences which are due to stretching/compressing of time and so can be expected to at least partly accommodate the effects of changes in download rate, queueing delay \etc.   

\begin{figure}
\centering
\includegraphics[width=0.8\columnwidth]{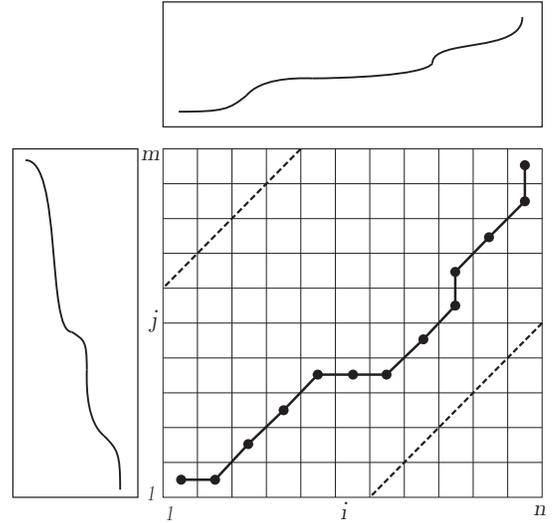}
\caption{Illustrating a warping path. The dashed lines indicate the warping window.}\label{fig:dtw}
\end{figure}

We define a warping path $\vv{p}$ to be a sequence of pairs, $\{(p^i_k,p^j_k)\}$, $k=1,2,\cdots,l$ with $(p^i_k,p^j_k)\in V:=\{1,\cdots,n\}\times\{1,\cdots,m\}$ satisfying boundary conditions $p^i_1=1=p^j_1$, $p^i_l=n$, $p^j_l=m$ and step-wise constraints $(p^i_{k+1},p^j_{k+1})\in V_{p^i_k,p^j_k}:=\{(u,v):u\in\{p^i_k,p^i_k+1\}\cap\{1,\dots,n\}, v\in\{p^j_k,p^j_k+1\}\cap\{1,\dots,m\}\}$, $k=1,\cdots,l-1$.   That is, a warping path maps points from one timestamp sequence to another such that the start and end points of the sequences match (due to the boundary conditions) and the points are monotonically increasing (due to the step-wise constraints).   This is illustrated schematically in Figure \ref{fig:dtw}, where the two timestamp sequences to be compared are indicated to the left and above the matrix and the bold line indicates an example warping path.   

Let $P^l_{mn}\subset V^l$ denote the set of all warping paths of length $l$ associated with two timestamp sequences of length $n$ and $m$ respectively, and let $C_{\vv{t},\vv{t}^\prime}(\cdot):P^l_{mn}\rightarrow\mathbb{R}$ be a cost function so that $C_{\vv{t},\vv{t}^\prime}(\vv{p})$ is the cost of warping path $\vv{p}\in P^l_{mn}$.  Our interest is in the minimum cost warping path, $\vv{p}^*(\vv{t},\vv{t}^\prime)\in\arg\min_{\vv{p}\in P^l_{mn}}C_{\vv{t},\vv{t}^\prime}(\vv{p})$.  In DTW the cost function has the separable form $C_{\vv{t},\vv{t}^\prime}(\vv{p})=\sum_{k=1}^l c_{\vv{t},\vv{t}^\prime}(p^i_k,p^j_k)$ where $c_{\vv{t},\vv{t}^\prime}:V\rightarrow\mathbb{R}$, in which case optimal path $\vv{p}^*(\vv{t},\vv{t}^\prime)$ be efficiently found using the backward recursion,
\begin{align}
&(p^i_k,p^j_k) \in \arg \min_{(p^i,p^j)\in V_k} C_{k+1} + c_{\vv{t},\vv{t}^\prime}(p^i,p^j) \label{eq:step}\\
&C_{k} = C_{k+1} + c_{\vv{t},\vv{t}^\prime}(p^i_k,p^j_k)
\end{align}     
where $V_k=(p^i,p^j)\in \{(u,v):(p^i_{k+1},p^j_{k+1})\in V_{u,v}\}$, $k=l-1,l-2,\cdots$ and  initial condition $C_l = c_{\vv{t},\vv{t}^\prime}(n,m)$.   When there is more than one optimal solution at step (\ref{eq:step}), we select $(p^i_k,p^j_k)$ uniformly at random from amongst them.

A common choice of element-wise cost is the Euclidean norm $c_{\vv{t},\vv{t}^\prime}(p^i,p^j)=(t_{p^i}-t_{p^j}^\prime)^2$.  However, in our data we found that this cost can lead to all the elements of one sequence that are beyond the last element of the other sequence being matched to that single element. For this reason and also to improve robustness to noise on the timestamp values (in addition to misalignment of their indices), following \cite{keogh01} we instead use the following element-wise cost
\begin{align}
c_{\vv{t},\vv{t}^\prime}(p^i,p^j) &= (D_{\vv{t}}(p^i) - D_{\vv{t}^\prime}(p^j))^2 
\end{align}
where $D_{\vv{t}}(i)=\frac{(t_i-t_{i^-})+(t_{i^+}-t_{i^-})}{2}$, $i^-=\max\{i-1,1\}$ and $i^+=\min\{i+1,|\vv{t}|\}$.  Observe that $D_{\vv{t}}(i)$  is akin to the derivative of sequence $\vv{t}$ at index $i$.  Further, we constrain the warping path to remain within windowing distance $w$ of the diagonal (\emph{i.e.} within the dashed lines indicated on Figure \ref{fig:dtw}) by setting $C(\vv{p})=+\infty$ for paths $\vv{p}\in P^l_{mn}$ for which $|p^i_k-p^j_k| > \max \{ w \min \{n , m \} , | m - n | \}$ for any $k\in\{1,\cdots,l\}$.   

\begin{figure}
\centering
\subfloat[Euclidean cost]{
  \includegraphics[width=0.45\columnwidth]{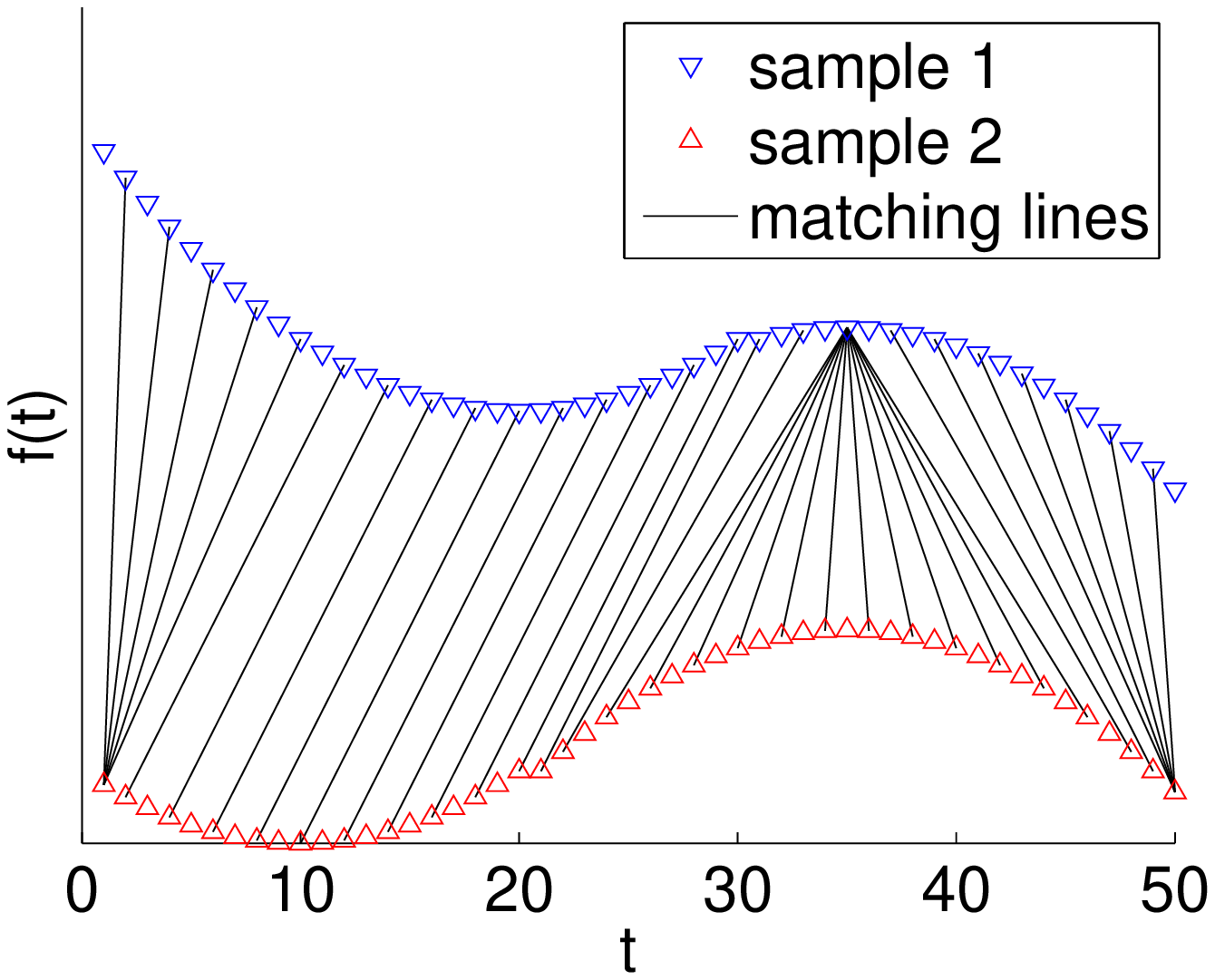}
  \includegraphics[width=0.45\columnwidth]{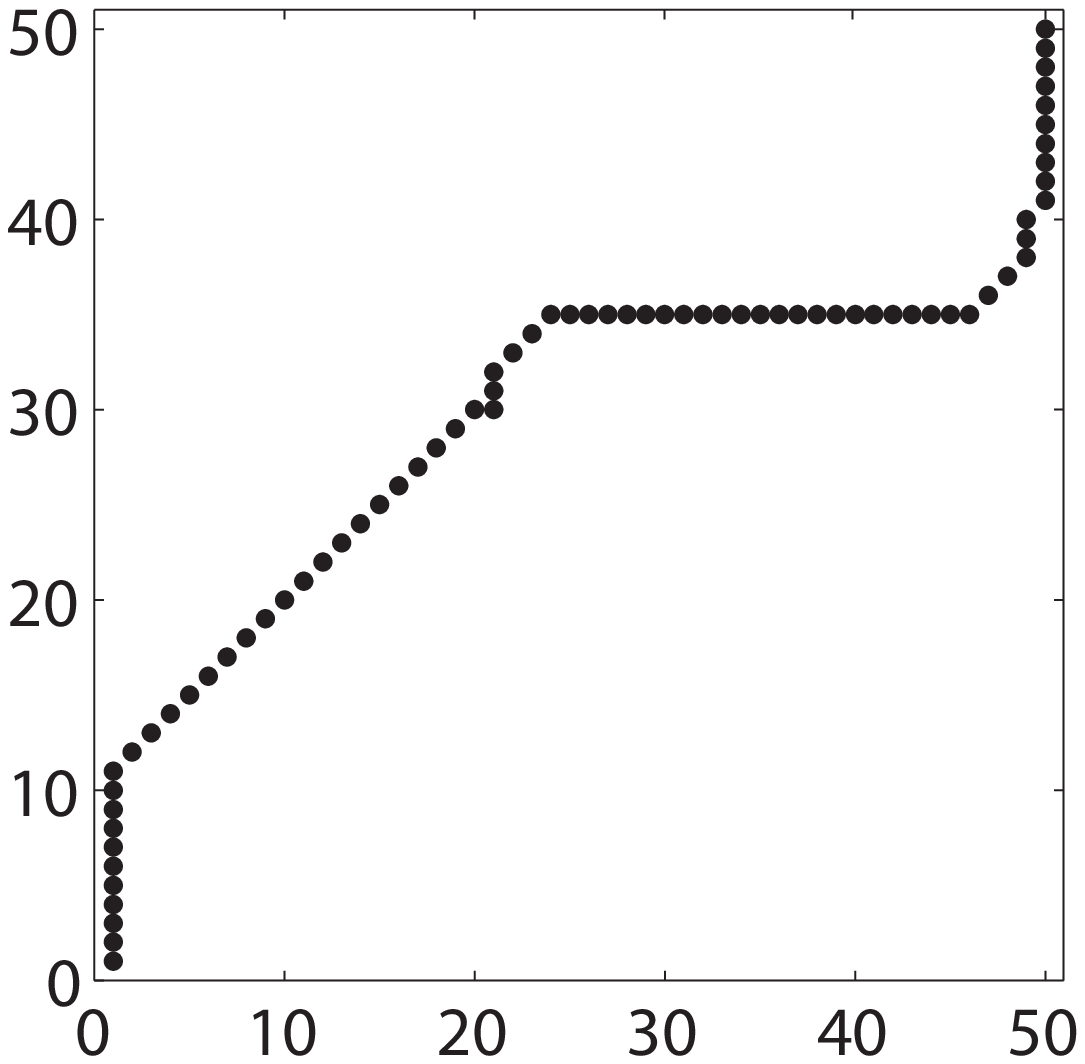}
	\label{fig:sdtwcomp}
}\\
\subfloat[Derivative cost]{
  \includegraphics[width=0.45\columnwidth]{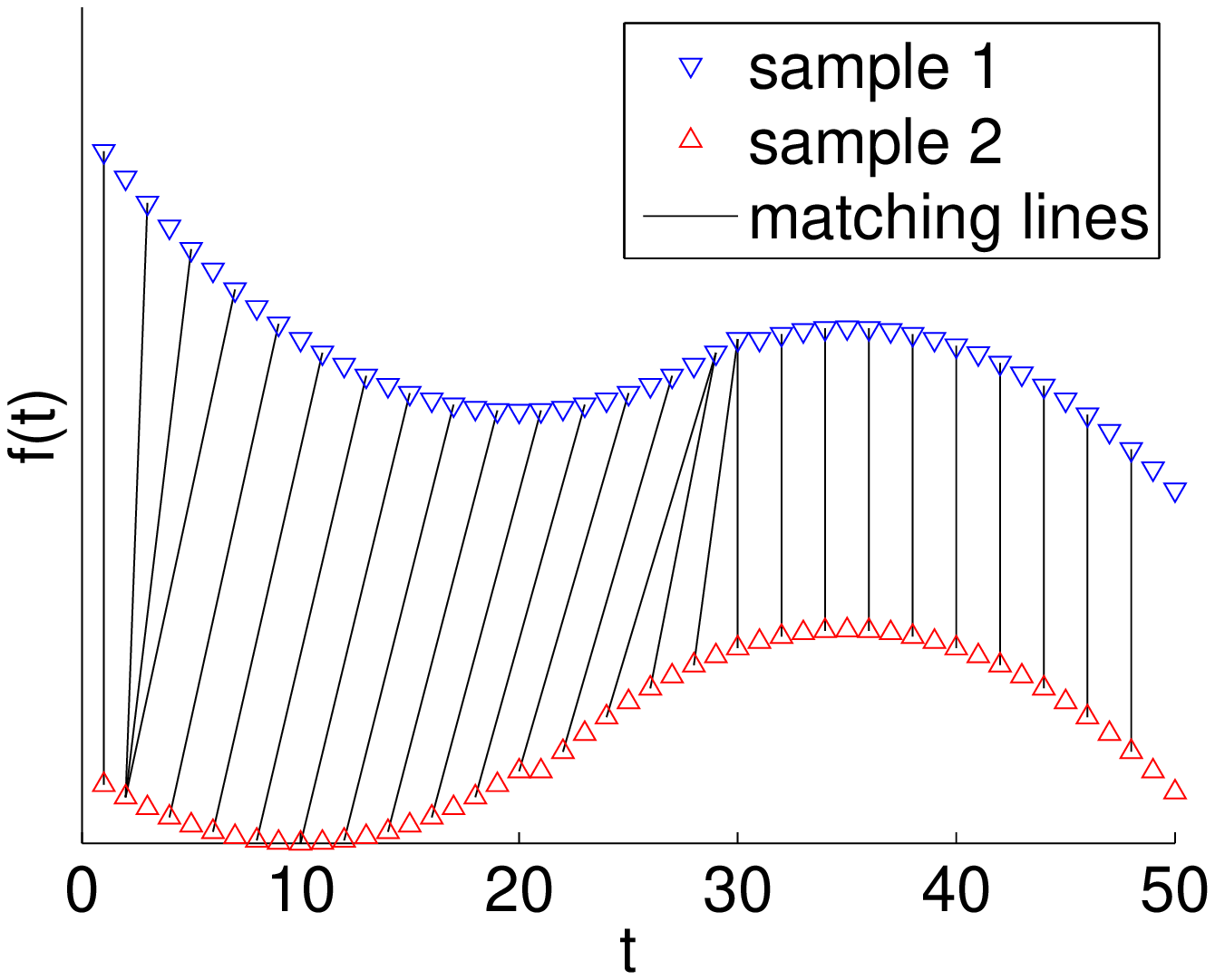}
  \includegraphics[width=0.45\columnwidth]{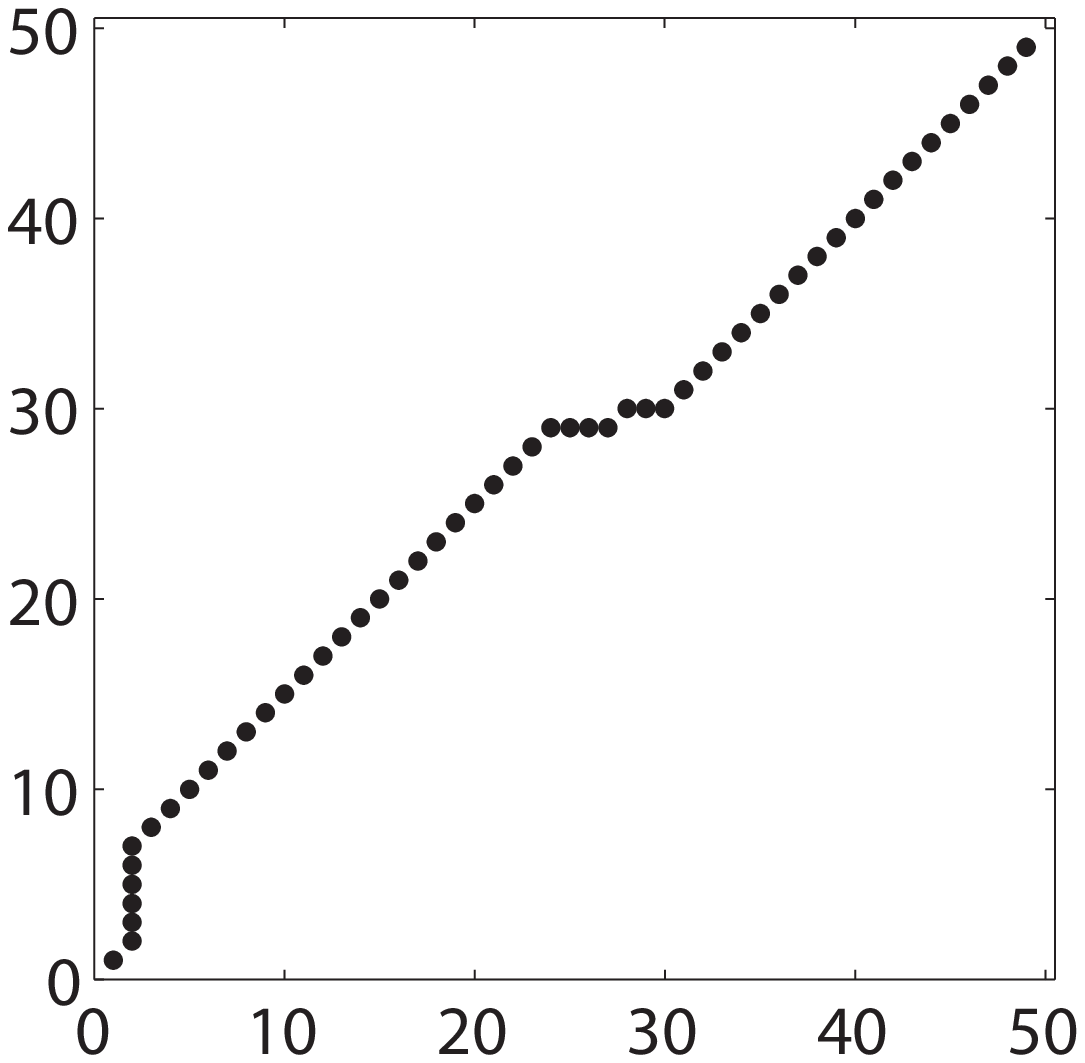}
	\label{fig:ddtwcomp}
}
\caption{Example DTW alignment and warping paths between two sequences vs cost function $c_{t,t^\prime}$ used, window $w=0.1$.  In this example the length $l$ of the warping path is 73 when a Euclidean cost is used and 54 with the derivative cost.}
\label{fig:derivative}
\end{figure}

Figure \ref{fig:ddtwcomp} illustrates the alignment of points between two sequences obtained using this approach and for comparison Figure \ref{fig:sdtwcomp} shows the corresponding result when using Euclidean cost.  The figure shows the warping paths on the right-hand side and an alternative visualisation of the mapping between points in the sequences on the left-hand side.   Observe that when Euclidean cost is used the warping path tends to assign many points on one curve to a single point on the other curve.  As noted in \cite{keogh01} this is known to be a feature of Euclidean cost.  In comparison, use of the derivative distance tends to mitigate this effect and select a warping path with fewer horizontal and vertical sections.

\subsection{$F$-Distance Measure}
\noindent
Given two timestamp sequences, the warping path is a mapping between them.   With reference to Figure \ref{fig:dtw}, sections of the warping path which lie parallel to the diagonal correspond to intervals over which the two sequences are well matched.   Sections of the warping path that are parallel to the x- or y-axes correspond to intervals over which the two sequences are poorly matched.   This suggests using the fraction of the overall warping path which is parallel to the  x- or y-axes as a distance measure, which we refer to as the $F$-distance.   

\begin{figure}
\centering
\subfloat[Sequence alignment.]{
  \includegraphics[width=0.4\columnwidth]{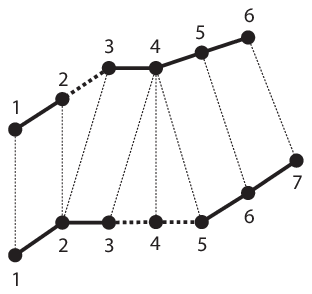}
	\label{fig:dtw_example}
}
\subfloat[Warping path]{
  \includegraphics[width=0.4\columnwidth]{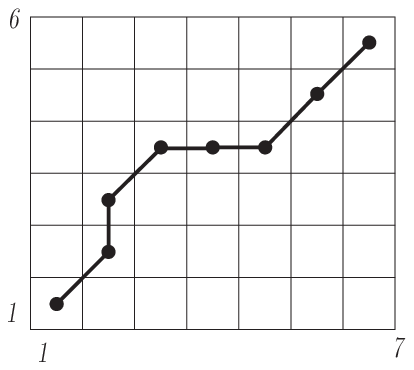}
	\label{fig:dtw_matching}
}
\caption{Illustrating method for calculating the $F$-distance between two timestamp sequences. }\label{fig:dtw_example2}
\end{figure}

In more detail, let $\vv{p}=\{(p^i_k,p^j_k)\}$, $k=1,\cdots,l$ be a derivative DTW warping path relating timestamp sequences $\vv{t}$ and $\vv{t}^\prime$, obtained as described in the previous section.   We partition the warping path into a sequence of subpaths within each of which either $p^i_k$ or $p^j_k$ remain constant and we count the subpaths which are longer than one.  For example, for the setup shown in Figure \ref{fig:dtw_example2} there are five subpaths: $(1,1)$; $(2,2), (2,3)$; $(3,4), (4,4),(5,4)$; $(6,5)$; $(7,6)$.   Two of these subpaths consist of more than one pair of points, namely $(2,2), (2,3)$ and $(3,4), (4,4),(5,4)$, and these correspond, respectively, to the vertical section and the horizontal section on the corresponding warping path shown in Figure \ref{fig:dtw_matching}.    

Formally, define $\kappa_1:=0<\kappa_2<\cdots<\kappa_{r-1}<\kappa_r:=l$ such that for each $s=1,\cdots,r-1$ (i) either $p^i_{k_1}=p^i_{k_2}$ $\forall k_1$, $k_2\in \{\kappa_s+1,\cdots,\kappa_{s+1}\}$  or $p^j_{k_1}=p^j_{k_2}$ $\forall k_1$, $k_2\in \{\kappa_s+1,\cdots,\kappa_{s+1}\}$ and (ii) either $\kappa_{s+1}=l$ or condition (i) is violated for some $k_1$, $k_2\in \{\kappa_s,\cdots,\kappa_{s+1}+1\}$ \emph{i.e.} each subsequence is maximal.   Note that $p^i_k\ne p^j_k$ for all $k=1,\cdots,l$ (due to warping path step-wise constraints) and so in condition (i) it is not possible for both $p^i_{k}$ and $p^j_{k}$ to be constant.   We are now is a position to define the $F$-distance measure between timestamp sequences $\vv{t}$ and $\vv{t}^\prime$, namely:
\begin{align}
\phi(\vv{t},\vv{t}^\prime) := \frac{\sum_{\substack{s \in \{1,\dots,r-1\} \\ \kappa_{s+1} - \kappa_s > 1}} \kappa_{s+1}-\kappa_s}{n + m}
\end{align}
where $\kappa_s$, $s=1,\cdots,r$ are the constant subsequences in minimal warping path $\vv{p}^*(\vv{t},\vv{t}^\prime)$.	It can be seen that $\phi(\vv{p})$ takes values in interval $[0,1]$, and is $0$ when sequences $\vv{t}$ and $\vv{t}^\prime$ are identical (in which case the warping path $\vv{p}$ lies on the diagonal in Figure \ref{fig:dtw}).  For the example in Figure \ref{fig:dtw_example2} the $F$-distance $\phi(\vv{p})$ is $(2+3)/13=0.385$.

\section{De-anonymising Web Fetches Over an Ethernet Tunnel}
\label{sec_basic}
\noindent
In this section we present measurements of web page queries carried out over an ethernet tunnel and evaluate the accuracy with which the web page being fetched can be inferred using only packet timing data. {The entire project including codes, scripts and datasets for all measurement campaigns is available at \cite{feghhi15}.}	The first dataset consists of home pages of each of the top Irish health, financial and legal web sites as ranked by www.alexa.com under its Regional/Europe/Ireland category in November 2014.  We prune the pages that fail to load and then for each of the top 100 sites we carry out 100 fetches of the index page yielding a total of 10,000 individual web page fetches in a dataset.  For comparison we collected two such datasets, one where the pages of each web site are fetched consecutively over an hour and a second where the pages are fetched each hour over a period of five days.   In these datasets the browser cache is flushed between each fetch so that the browser always starts in a fresh state.     In addition, a third dataset was collected consisting of the same 10,000 web fetches but now without flushing of the browser cache between fetches.  The web pages were fetched over a period spanning November 2014 to January 2015.   A \texttt{watir-webdriver} script on Firefox 36.0 was used to perform the web page fetches and \emph{tcpdump} to record the timestamps and direction (uplink/downlink) of all packets traversing the tunnel although only packet timestamps on the uplink were actually used. 

\subsection{Hardware/Software Setup}
\noindent
The network setup consists of a client that routes traffic to the internet over a gigabit ethernet LAN. The client machine is a Sony VGN-Z11MN laptop with an Intel core 2 duo 2.26GHz CPU and 4GB of memory. It is running Ubuntu Linux 14.04 LTS Precise.

\subsection{Classifying Measured Timestamp Sequences}
\noindent
We use the $F$-distance measure $\phi(\cdot,\cdot)$ described in Section \ref{sec:distance} to compare measured uplink timestamp sequences, with windowing parameter $w=0.2$ unless otherwise stated.  

\begin{figure}
\centering
\subfloat[]{
  \includegraphics[trim=1cm 0cm 1cm 0cm,clip, width=0.5\columnwidth]{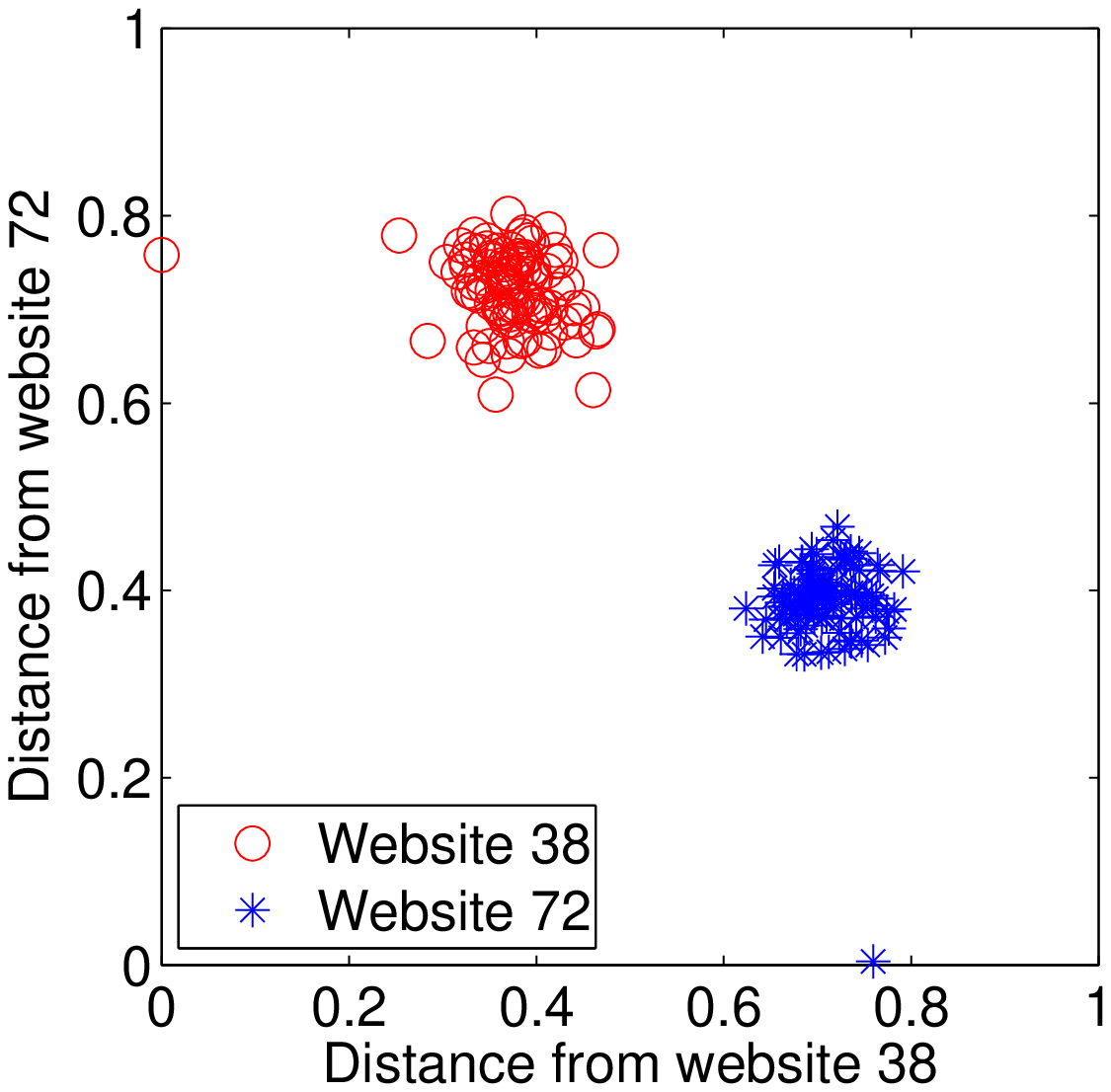}
	\label{fig_distinct}
}
\subfloat[]{
  \includegraphics[trim=1cm 0cm 1cm 0cm,clip, width=0.5\columnwidth]{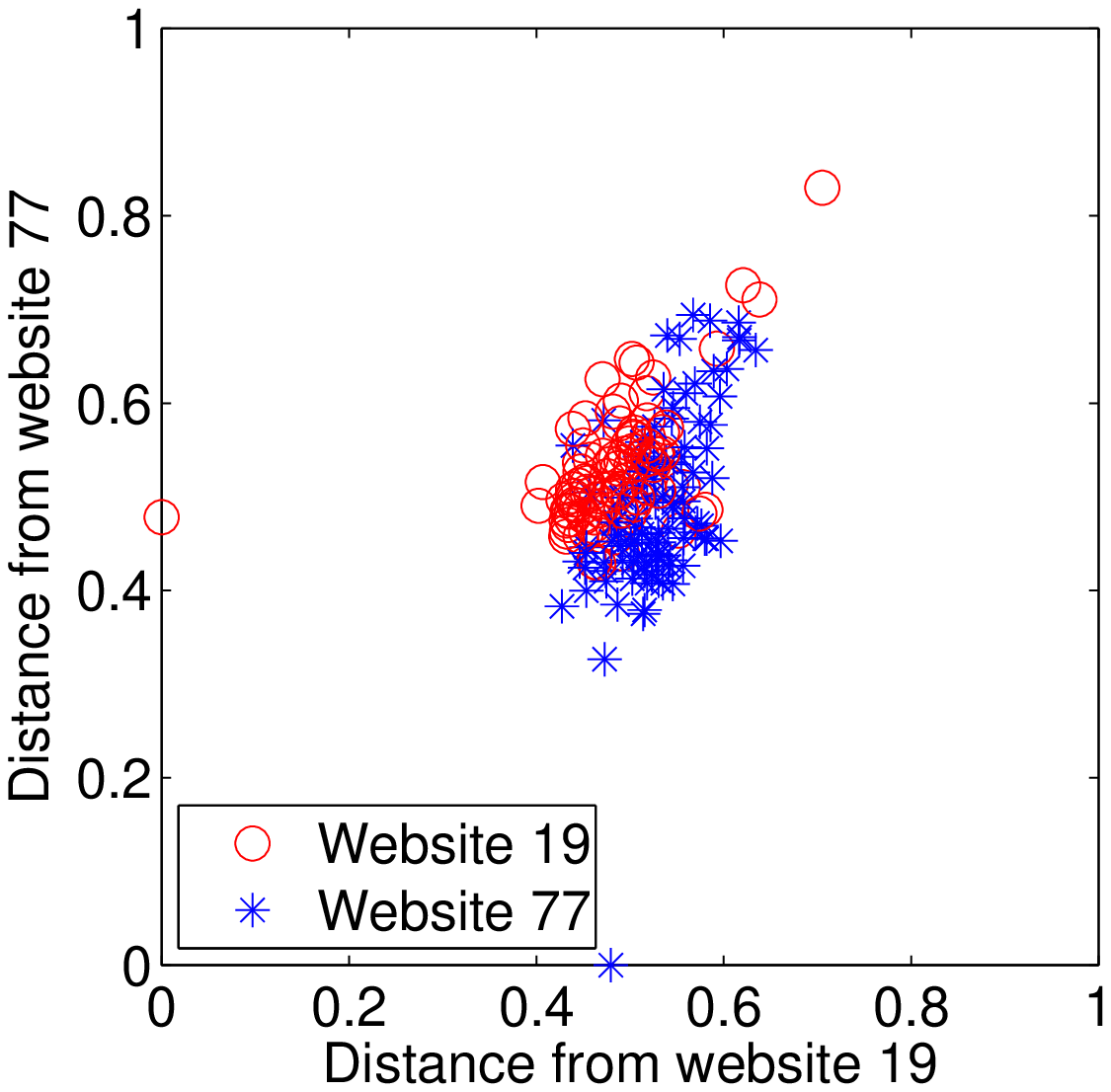}
	\label{fig_close}
}
\caption{Scatter plots for 4 different web pages using $F$-distance measure $\phi$. In (a) two relatively distinct web pages are compared while the web pages in (b) are relatively similar.  }\label{fig:scatter}
\end{figure}

Figure \ref{fig:scatter} shows example scatter plots obtained using this distance measure.   In more detail, from the set $T_i$ of measured timestamp sequences for the $i$-th web site we select a sequence $\vv{t}_i$ which minimises $\sum_{\vv{t}\in T_i}\phi(\vv{t},\vv{t}_i)$ and then use $\vv{t}_i$ as the exemplar for the $i$-th web page.  In Figure \ref{fig:scatter} we then plot $\phi(\vv{t},\vv{t}_i)$ for each of the timestamp sequences $\vv{t}$ measured for web page $i$ and also for timestamp sequences measured for another web page.  In the example in Figure \ref{fig_distinct} it can be seen that the distance measure is indeed effective at separating the measured timestamp sequences of the two web pages considered into distinct clusters, so potentially providing a basis for accurately classifying timestamp sequences by web page.   Figure \ref{fig_close} shows an example of a scatter plot where the separation between the two web pages is less distinct and so classification can be expected to be less reliable.  As we will see, examples of this latter sort turn out to be fairly rare.

We considered two approaches for using $\phi(\cdot,\cdot)$ to classify timestamp sequences: $K$-Nearest Neighbours and Naive Bayes Classification.

\subsubsection{K-Nearest Neighbours}\label{sec:knn}
\noindent
In this method, for each web page $i$ we sort the measured timestamp sequences $\vv{t}^\prime\in T_i$ used for training in ascending order of sum-distance $\sum_{\vv{t}\in T_i}\phi(\vv{t},\vv{t}^\prime)$ and select the top $3$ to use as exemplars to represent this web page.   When presented with a new timestamp sequence, its distance to the exemplars for all of the training web pages is calculated and these distances are sorted in ascending order.   Classification is then carried out by majority vote amongst the top $K$ matches.

\begin{figure*}
\centering
\subfloat[$K=1$]{
  \includegraphics[trim=0cm 1cm 0cm 1cm,clip,width=0.33\textwidth]{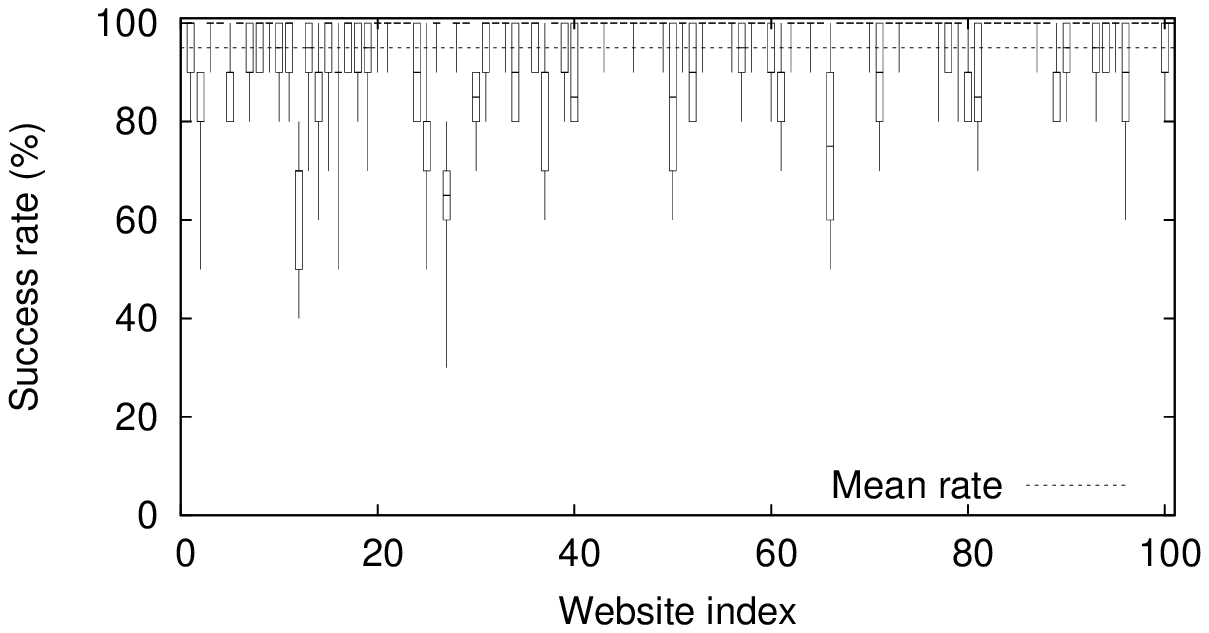}
}
\subfloat[$K = 3$]{
  \includegraphics[trim=0cm 1cm 0cm 1cm,clip,width=0.33\textwidth]{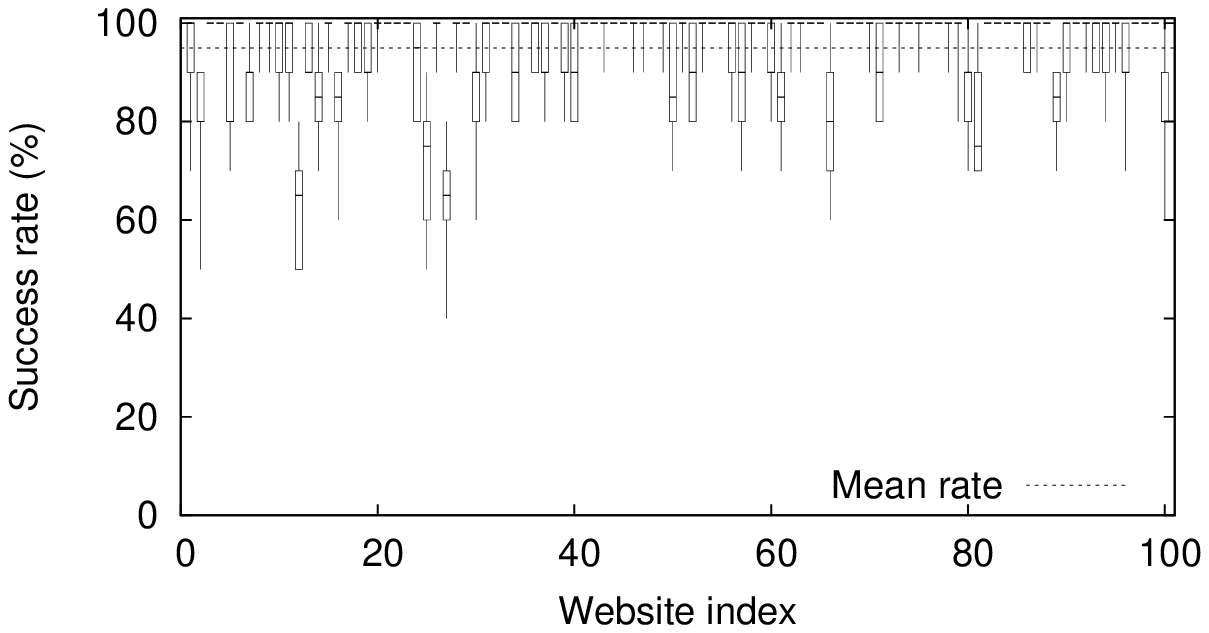}
}
\subfloat[$K = 5$]{
  \includegraphics[trim=0cm 1cm 0cm 1cm,clip,width=0.33\textwidth]{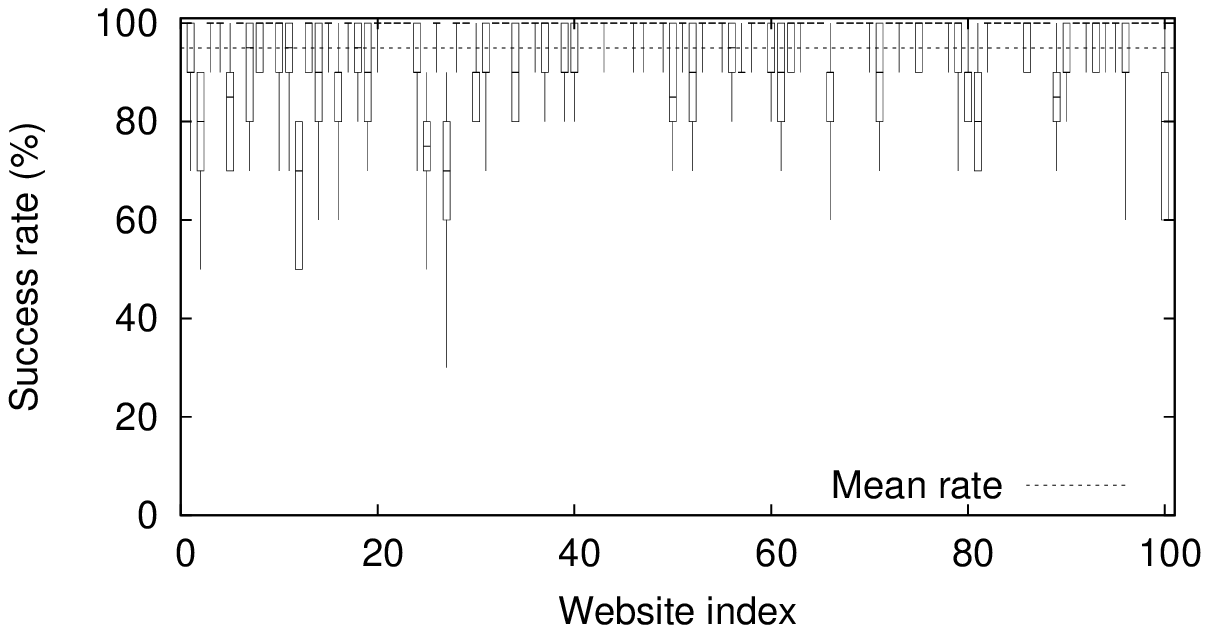}
}
\caption{$K$-Nearest Neighbours classification performance, no browser caching. }
\label{fig_normal_40}
\end{figure*}

\subsubsection{Naive Bayes Classifier}
\noindent
For each web page $i$ from the measured timestamp sequences $T_i$ used for training we select $\vv{t}_i\in \arg\min_{\vv{t}^\prime\in T_i}\sum_{\vv{t}\in T_i}\phi(\vv{t},\vv{t}^\prime)$ (in addition we also consider selecting $\vv{t}_i$ to minimise the variance of the distance $\phi$, see below) and then fit a Beta distribution to the empirical distribution of $\phi(\vv{t},\vv{t}_i)$ for $\vv{t}\in T_i$.   Let $p_i(\cdot)$ denote the probability distribution obtained in this way.  When presented with a new timestamp sequence $\vv{t}$, we calculate the probability $p_i(\vv{t})$ of this sequence belonging to web page $i$ and select the web page for which this probability is greatest.

\begin{figure}[!htb]
\subfloat[Minimum Mean]{
  \includegraphics[trim=0cm 1cm 0cm 1cm,clip,width=1.0\columnwidth]{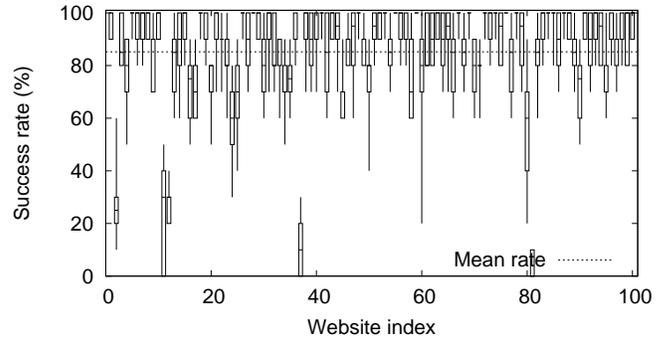}
	\label{fig_bayes_mean}
}\\
\subfloat[Minimum Variance]{
  \includegraphics[trim=0cm 1cm 0cm 1cm,clip,width=1.0\columnwidth]{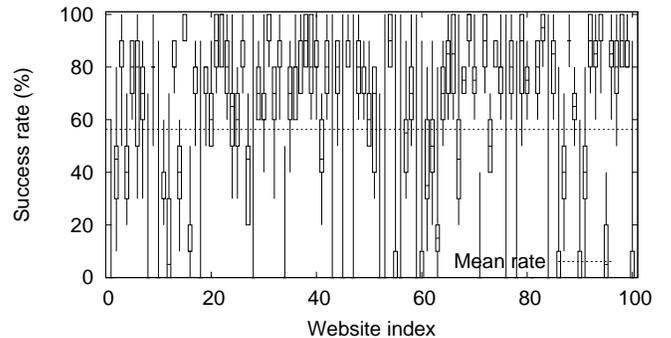}
	\label{fig_bayes_var}
}
\caption{Naive Bayes classification performance, no browser caching.}\label{fig:bayes}
\end{figure}

\subsection{Experimental Results}\label{sec:ether}
\noindent
We begin by presenting results for the dataset where pages are fetched consecutively and the browser cache is flushed between fetches.   Figure \ref{fig_normal_40} details the measured classification accuracy using the $K$-NN approach, for various values of $K$.  We use $10$-fold cross validation, where the 100 samples of each web site are divided into 10 random subsets and for each subset we use the remaining 90 samples as the training data to find the exemplars and use the 10 samples in the subset as the validation data.  The rates for these 10 subsets for each web site are summarized and displayed in the figure. Each of the boxes indicate the $25\%$, $50\%$ and $75\%$ quartiles and the lines indicate the maximum and minimum values. The mean success rates for $K=1$, $K=3$ and $K=5$ are $95.01\%$, $94.97\%$ and $94.98\%$ respectively. {These results for uplink traffic compares to a maximum success rate of $92.5\%$ when using packet timestamps on the downlink for the classification, indicating that use of uplink or downlink timestamps has little effect on the performance of this classification attack.}	The results are also compared for a subset of $50$ web sites selected randomly from the current $100$, see Table \ref{tab:knn}, which also confirms that the effect of population size is minor.

For comparison, the success rates when web pages are fetched hourly over 5 days are $90.88\%$, $90.72\%$ and $90.74\%$.  Observe that there is a small (about 5\%) reduction in success rate, which we assume is associated with the time-varying nature of some of the web sites.	{We discuss the effect of content and speed variability on the performance in \mbox{Section \ref{sec:content_change}.}}

Figure \ref{fig:bayes} plots the corresponding results obtained using the naive Bayes approach.  Performance is calculated when the exemplar for each web page is selected to minimise the mean and the variance of the distance.	The mean success rates are $85.2 \%$ and $56.3 \%$ respectively.  Since the performance is consistently worse than that of the $K$-NN classifier we do not consider the naive Bayes approach further in the rest of the paper.

\subsection{Standard vs. Cached: Different Versions of Same Web Page}
\noindent
On first visiting a new web page a browser requests all of the objects that form the web page.   However, on subsequent visits many objects may be cached \emph{e.g.} images, css and js files, \emph{etc}.  In the Mozilla browser, when the address of a web page is simply entered again shortly after the full page is fetched, since the cached copy of an object has not yet expired the cached copy will be used when rendering the web page and it will not be fetched over the network by the browser. But the browser can be forced to reload the web page by pressing F5 where it then sends a request for the objects and the server may either return an abbreviated NOT MODIFIED response if the cached object is in fact still fresh or return the full object if it has changed. Ultimately a full refresh can be induced by pressing Ctrl+F5 which requests for the full version of the web page as if no object is cached before.  Hence, the network traffic generated by a visit to a web page may differ considerably depending on whether it has been visited recently (so the cache is fresh) or not.

\begin{figure}
\centering
\includegraphics[trim=0cm 1cm 0cm 1cm,clip,width=0.5\textwidth]{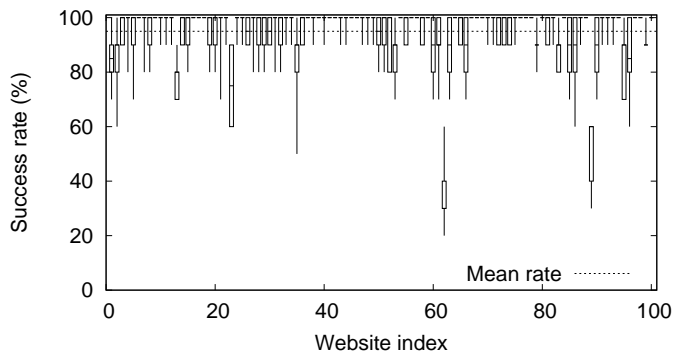}
\caption{$K$-Nearest Neighbour classification performance, with browser caching using $3$ exemplars for each site.  $K=5$.}\label{fig:cached}
\end{figure}

Classification of cached web pages can be expected to be more challenging than for non-cached pages since there is less network traffic and so less data upon which to base the classification decision.  Figure \ref{fig:cached} presents the measured classification accuracy when browser caching is enabled.   This data is for the case where requests that reply with NOT MODIFIED use the cached content, which is probably the most common form of caching used in practice. It can be seen that regardless of the small size of the network traffic in this setup,   the overall success rate for identifying web pages remains in excess of $95\%$.

\subsection{Web Pages Outside the Training Set}\label{sec:falsepos}
\noindent
The experiments in the previous two sections are conducted with the assumption that the adversary knows that the web page that the user has visited is among the set of web pages for which training data has been collected.
When this assumption need not hold, \emph{i.e.} the user might have fetched a web page outside of the adversary's training database, then we can use the following approach to first classify whether a measured packet timestamp sequence $\vv{t}$ is associated with a web site in the training set or not.

Recall that, as discussed in Section \ref{sec:knn}, for each web page $i$ in the training set we have $3$ exemplar packet timestamp sequences that are used for $K$-Nearest Neighbour classification.	   Given a packet timestamp sequence $\vv{t}$ we use $K$-Nearest Neighbour classification to estimate the nearest web page $w(\vv{t})$ within the training set and let $F_{min}(\vv{t})$ denote the minimum $F$-distance between the exemplars for this web page and the measured timestamp sequence.   We can then use this value as the basis for a simple classifier.  Namely, when $F_{min}(\vv{t})$ is greater than a specified threshold (which may depend on $w(\vv{t})$) then we estimate $\vv{t}$ as lying outside the training set, and when $F_{min}(\vv{t})$ is below the threshold then we estimate $\vv{t}$ as lying within the training set.  It remains to select an appropriate threshold for each web page in the training set.

\begin{figure}
\centering
\includegraphics[trim=0cm 1cm 0cm 1cm,clip,width=0.5\textwidth]{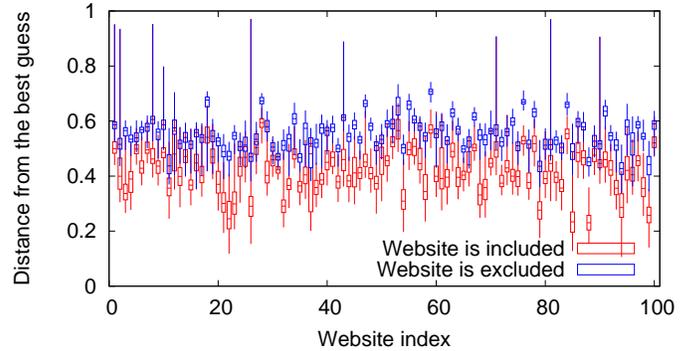}
\caption{Distribution of the $F$-distance between the measured packet timestamp sequences in the training dataset and the exemplar packet sequences for the best guess. Data is shown for when sequences of each web site are within the training dataset and for when they are removed.  Ethernet channel, no browser caching.}\label{fig:hamilton_dist}
\end{figure}

For every timestamp sequence $\vv{t}$ in the training set Figure \ref{fig:hamilton_dist} plots the distribution of $F_{min}(\vv{t})$ vs the index of the web site for which $\vv{t}$ is measured.   This figure is a box and whiskers plot with the min, max and quartiles shown.   For every web site we then remove its data from the training set and repeat the calculation.  The distribution of these values is also shown in Figure \ref{fig:hamilton_dist}.  It can be seen that, unsurprisingly, the $F$-distance is consistently higher when a web site is excluded from the training set.   We select the threshold for classification to try to separate these two sets of value.   Namely, we take the average of the $x$ percentile of the lower values and the $(100-x)$ percentile of the upper values as our threshold,  where $0\le x \le 100$ is a design parameter.

\begin{figure}
\centering
\subfloat[Mean]{
  \includegraphics[width=0.5\columnwidth]{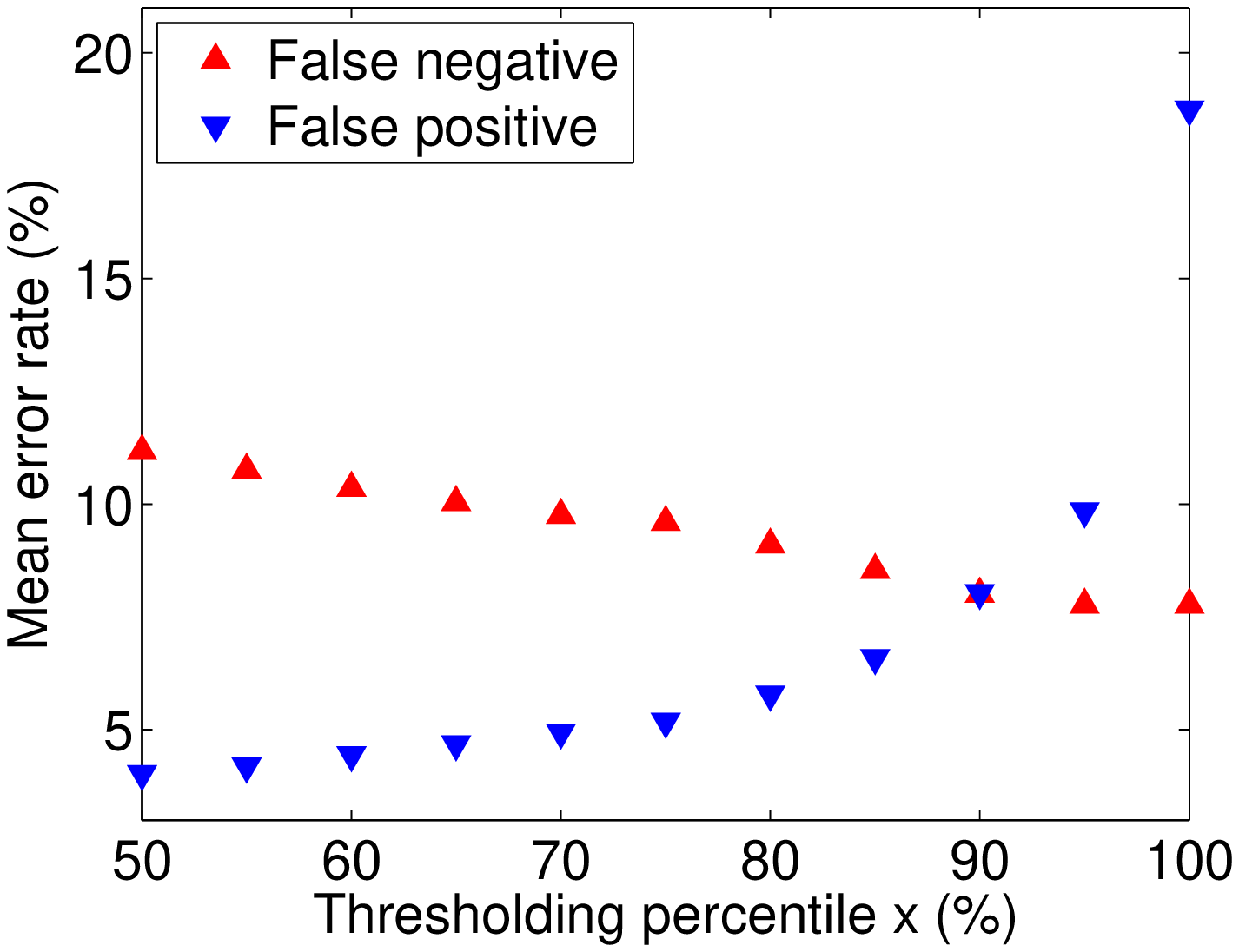}
	\label{fig:thr_meanerr}
}
\subfloat[Standard Deviation]{
  \includegraphics[width=0.5\columnwidth]{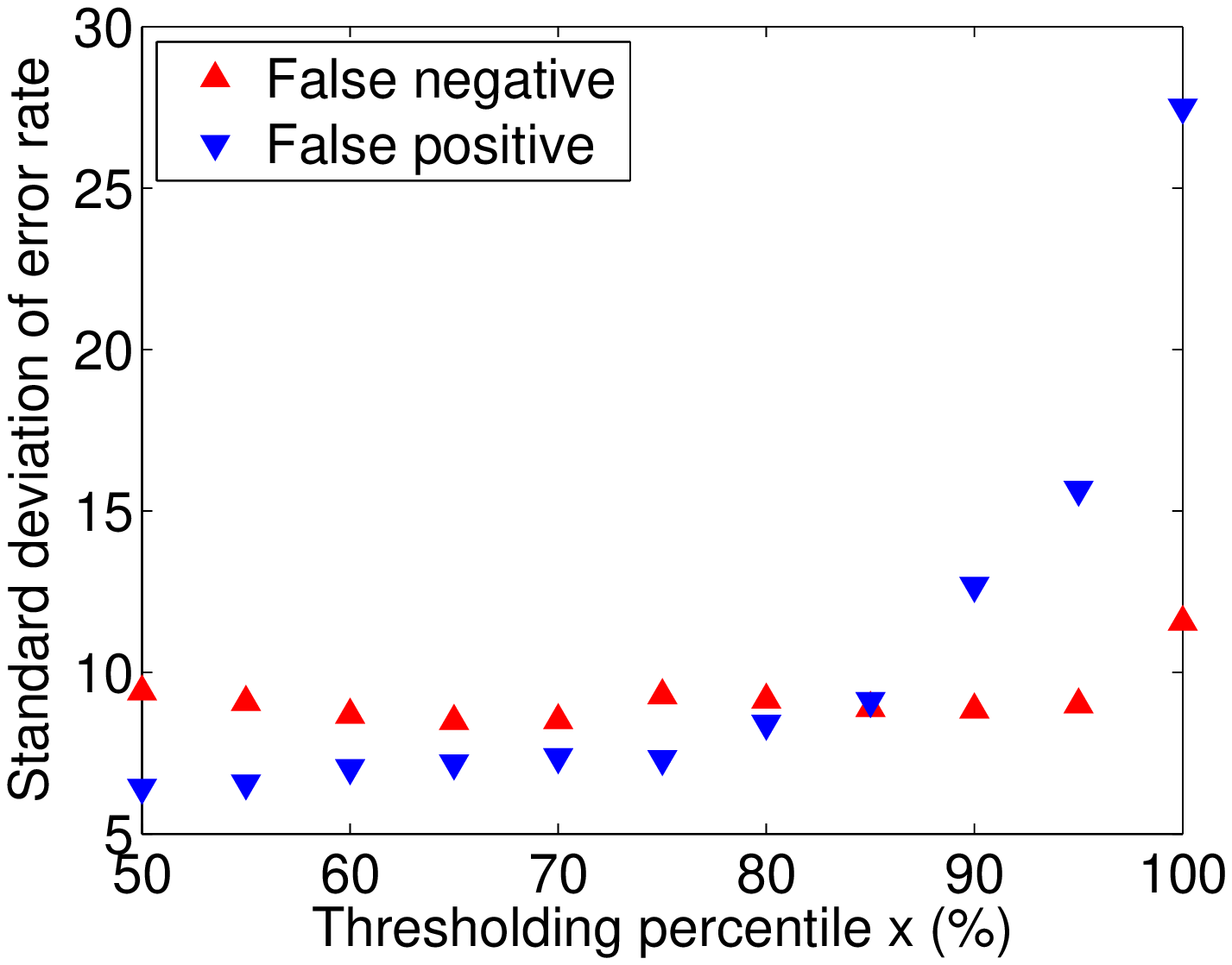}
	\label{fig:thr_stderr}
}
\caption{Mean and standard deviation of false negative and false positive error rates vs the choice of $F$-distance threshold (specified via design parameter $x$).}
\end{figure}

The classification error rate vs the threshold parameter $x$ used is shown in Figure \ref{fig:thr_meanerr}.  Two error rates are shown, firstly the fraction of web pages which are outwith the training set but which are classified as lying within it (which we refer to in this section as false positives) and secondly the fraction of web pages which are within the training set but which are classified as lying outwith it (which we refer to as false negatives).   The standard deviations of these error rates across the web pages is also shown in Figure \ref{fig:thr_stderr}.  It can be seen that thresholding with $x=90$ yields equal error false negative and false positive rates of about $8.0\%$, which is close to the complement of the reported success rate reported in the preceding section.

\section{Measurement Results for Other Channels}
\noindent
In this section we extend consideration from ethernet to a number of different network channels.  Namely, we consider packet timestamp measurements taken from a commercial femtocell carrying cellular wireless traffic, from a time-slotted wired UDP channel (of interest as a potential defence against timing analysis) and from the first hop (\emph{i.e.} between the client and the Tor gateway) of a Tor channel.   Similar to before, in each case we collected packet timestamp data for 100 fetches of the home pages of each of the top 100 Irish health, financial and legal web sites as ranked by www.alexa.com.

\subsection{Femtocell Traffic}
\noindent
A femtocell is an eNodeB cellular base station with a small physical footprint (similar to a WiFi access point) and limited cell size (typically about 30m radius).  It is intended to improve cellular coverage indoors, filling in coverage holes and improving download rates, while also offloading traffic from the macrocell network.  Wired backhaul to the cellular operators network is via a user supplied network connection \emph{e.g.} a home DSL line.   Since femtocells are usually user installed, physical access to the backhaul connection is straightforward and it is a simple matter to route backhaul traffic via a sniffer.  Mobile operators are, of course, aware of this and backhaul traffic is therefore secured via use of an IPSec encrypted tunnel.   In the setup considered here, the femtocell backhaul is over a university gigabit ethernet connection and we used \emph{tcpdump} to log packets passing over this link.  

\subsubsection{Hardware/Software Setup}
\noindent
The client computer is the same Sony laptop used for the ethernet measurements. It now uses a Huawei K3770 HSPA USB Broadband Dongle to connect wirelessly to the internet via a Femotcell. The femtocell is a commercial Alcatel-Lucent 9361 Home Cell V2-V device.  The femtocell wired backhaul is connected to a campus network via a NetGear EN 108 TP Ethernet hub.   A monitor computer which is running on a AMD Athlone 64 X2 Dual Core Proc 5000+ CPU and 4GB memory is also connected to this hub and logs all packets.   The client and monitor computers both run Ubuntu Linux 14.04 LTS Precise.

\subsubsection{Results}
\begin{figure}
\centering
 \includegraphics[trim=0cm 1cm 0cm 1cm,clip,width=0.50\textwidth]{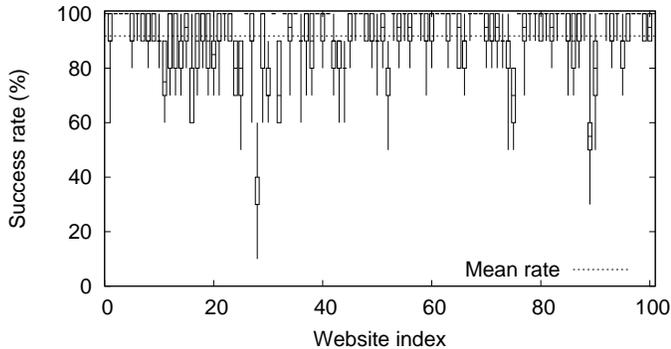}
\caption{Femtocell $K$-Nearest Neighbours classification performance, no browser caching, $K = 5$.}
\label{fig:femto-comp}
\end{figure}

\noindent
In contrast to the relatively clean ethernet channel considered in Section \ref{sec:ether}, we found that traffic passing over the wireless femtocell link is often distorted by factors such as wireless and cellular noise, encoding/decoding delays, cellular control plane traffic \etc.	These distortions typically appear as shifts along the $x$-axis of the packet timestamp patterns and/or as delays in the $y$-axis. The measured performance using a $K$-NN classifier using $3$ exemplars for each site and $K=5$ is shown in Figure \ref{fig:femto-comp}.  The mean success rate is $91.8\%$, which compares with the mean success rate of $95\%$ observed in Section \ref{sec:ether} when using a clean ethernet channel.  It can be seen that use of the wireless channel tends to reduce the classification accuracy, as might be expected due to the additional loss/delay over the wireless hop.  However, the reduction in accuracy is minor.

\begin{figure}
\centering
\subfloat[Slot size: 1ms]{
  \includegraphics[trim=0cm 1cm 0cm 1cm,clip,width=0.5\textwidth]{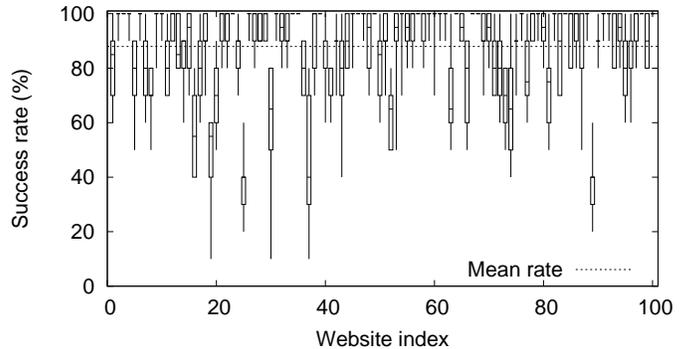}
	\label{fig:slotted-1}
}\\
\subfloat[Slot size: 10ms]{
  \includegraphics[trim=0cm 1cm 0cm 1cm,clip,width=0.5\textwidth]{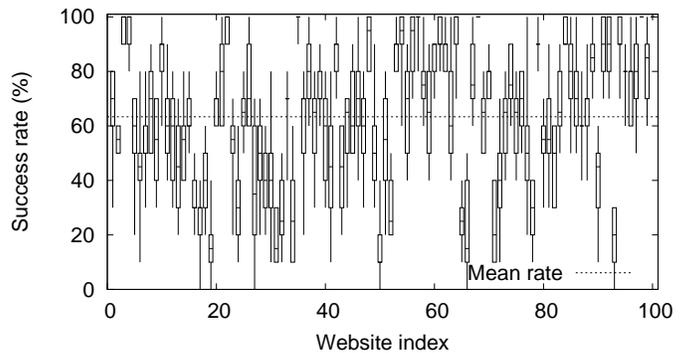}
	\label{fig:slotted-10}
}
\caption{Time-slotted tunnel $K$-Nearest Neighbours classification performance,  no browser caching, $K = 5$.}\label{fig:slotted}
\end{figure}

\subsection{Time Slotted UDP Tunnel}
\noindent
We developed a custom tunnel using \texttt{iptables}, \texttt{netfilter} and \texttt{netfilter-queue}.  The tunnel transports packets over a UDP channel in a time slotted fashion and the slot size is a configurable parameter.  

\subsubsection{Hardware/Software Setup}
\noindent
The experimental setup is identical to that used in Section \ref{sec_basic} apart from the use of a customised tunnel.   On the client computer all web traffic is captured using the OUTPUT \texttt{netfilter} hook, encapsulated into UDP packets and sent to a server at the other side of the tunnel.  The server, which has an AMD Athlone 64 X2 Dual Core Proc 5000+ and 4GB memory, fetches these UDP packets using the PREROUTING hook, extracts the payload and sends them by via the FORWARD hook to the outgoing ethernet interface.  Similarly, incoming packets from the internet are encapsulated into UDP packets via FORWARD hook on the server and sent to the client which captures them using the PREROUTING hook, extracts the payload and forwards this to the application layer.

\subsubsection{Results}
\noindent
Figure \ref{fig:slotted} shows the measured performance using a $K$-NN classifier where $3$ exemplars are chosen from each site and $K=5$.  The overall success rate is $88\%$ when the tunnel slot size is $1$ms and $63\%$ when the tunnel slot size is increased to $10$ms.  We also considered slot sizes larger than 10ms, but since we found such that large slot sizes tended adversely affect browser performance (and so would likely be problematic in practice) we do not include them here.  This performance compares with a success rate of $95\%$ over a plain ethernet tunnel.   As might be expected, time-slotting decreases the classification success rate since it adds timing ``noise''.   However, even with a relatively large slot size of 10ms the impact on performance is not proportional to the sacrifice we make in terms of delay and throughput (with such a large slot size we are capping the downlink throughput to $150$KB/s).  This approach therefore appears to be unappealing as a practical defence against the timing-based attack considered here.  Of~course more sophisticated types of defence may be more effective, but we leave consideration of those to future work as they likely involve complex trade-offs between network performance and resistance to attack that we lack space to address here.

\subsection{Tor Network}
\noindent
In this section we consider measurements of web page queries over the Tor network.   Tor is an overlay network of tunnels that aims to improve privacy and security on the internet.  

\subsubsection{Hardware/Software Setup}
\noindent
The experimental setup is the same as in Section \ref{sec_basic} except that the traffic from the client browser, Mozilla Firefox 36.0 is proxified over Tor v0.2.5.11.  Note that we also explored use of the Tor browser but found that a significant subset of the web sites failed to load, timed out or required a CAPTCHA to be solved for each page fetch which created complications when scripting fetches.  We also investigated using Firefox with Tor pluggable transports (such as obfs4 \etc.) but we found that using these add-ons had a huge impact on delay such that most web sites fail to load even after 5 minutes.	As before, the browser cache is flushed between fetches.

\subsubsection{Randomised Routing}
\noindent
Tor uses randomised routing of traffic over its overlay network in an attempt to make linking of network activity between source and destination more difficult.   It can be expected that rerouting will have a significant impact on the timestamp sequence measured during a web fetch since changes in path propagation have a direct impact on the time between an outgoing request and receipt of the corresponding server response, and also impact TCP dynamics since congestion window growth slows with increasing RTT.  Differences in loss rate, queueing delay \etc. along different routes are also likely to impact measured timestamp sequences.  

\begin{figure}[!htb]
\centering
\subfloat[Mean RTT for packets of each sample]{
  \includegraphics[trim=0cm 1.5cm 0cm 2cm,clip, width=0.8\columnwidth]{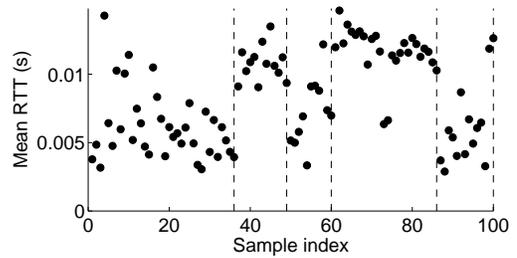}
	\label{fig:mean-rtt}
}\\
\subfloat[Max RTT for packets of each sample]{
  \includegraphics[trim=0cm 1.5cm 0cm 2cm,clip, width=0.8\columnwidth]{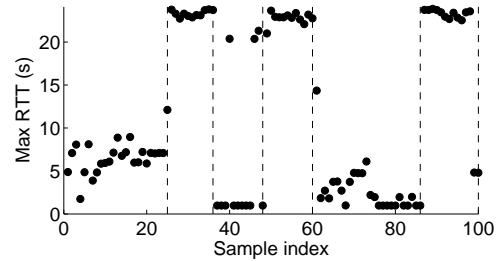}
	\label{fig:max-rtt}
}
\caption{Mean and max RTTs measured during 100 fetches of the web page www.medicalcouncil.ie.  Changes due to Tor rerouting are evident.   The max RTT in (b) is in fact the idle time between when the last packet is received until the browser is closed, hence why it is significantly larger than the mean RTT plotted in (a).}\label{fig:rerouting}
\end{figure}

The impact of Tor rerouting on measured RTT is illustrated in Figure \ref{fig:rerouting}, which plots the mean and max delay between sending of a TCP data packet and receipt of the corresponding TCP ACK for repeated fetches of the same web page (although this information is not available to an attacker, in our tests it is of course available for validation purposes).   Abrupt, substantial changes in the mean RTT are evident, especially in Figure \ref{fig:rerouting}b.  These changes persist for a period of time as Tor only performs rerouting periodically.

\begin{figure}[!htb]
\centering
	\includegraphics[trim=0cm 0cm 0cm 1.7cm,clip, width=0.8\columnwidth]{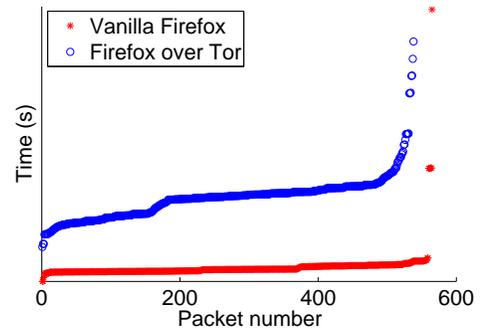}
	\caption{Time traces of uplink traffic measured when fetching www.medicalcouncil.ie .  Measurements are shown both when using vanilla Firefox and when using Firefox with the Tor plugin.}
	\label{fig:diftor}
\end{figure}

Figure \ref{fig:diftor} illustrates the impact of Tor on the packet timestamps measured during a web page fetch. 

\begin{figure}[!htb]
\centering
  \includegraphics[trim=0cm 1cm 0cm 1cm,clip,width=0.5\textwidth]{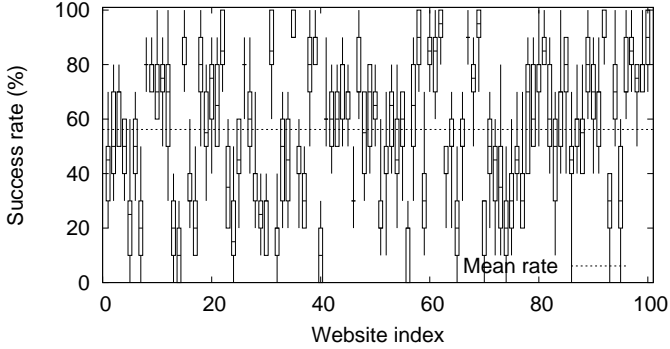}
\caption{Tor network $K$-Nearest Neighbours classification performance,  no browser caching, $K = 5$.}\label{fig:tor-100}
\end{figure}

\subsubsection{Results}\label{tor:results}
\noindent
Figure \ref{fig:tor-100} details the measured classification accuracy using the $K$-NN approach, where $3$ exemplars are chosen from each site and a window size of $w = 0.2$ is used to accommodate the warping between samples.   The mean success rate is $56.2\%$ which compares with the mean success rate of $95.0\%$ when using a clean ethernet channel.  As might be expected, use of the Tor network significantly reduces classification accuracy.  However, the success rate of $56.2\%$ compares with a baseline success rate of $1\%$ for a random classifier over 100 web sites and so still is likely to represent a significant compromise in privacy.   We note also that this compares favourable with the $54.6\%$ rate reported by Panchenko \emph{et al} in \cite{panchenko11} against Tor traffic using packet size and direction information.

\begin{table*}[!htb]
\centering
	\begin{tabular}{| c | c | c | c | r | r | r | r | }
		\hline
		\multicolumn{2}{|c|}{\multirow{2}{*}{Channel}} & \multirow{2}{*}{\specialcell{Number of\\Exemplars}} & \multirow{2}{*}{\specialcell{Database\\size}} & \multicolumn{4}{c |}{K} \\ \cline{5-8}
		\multicolumn{2}{|c|}{} & & & 1 & 3 & 5 & 7 \\ \hline
		\multicolumn{2}{|c|}{\multirow{5}{*}{Ethernet}} & 5 & 100  & 95.27\% & 95.65\% & 95.86\% & 95.74\% \\
		 \multicolumn{2}{|c|}{} & 3 & 100  & 95.01\% & 94.97\% & 94.98\% & - \\
		 \multicolumn{2}{|c|}{} & 3 & 100$^*$ & 90.88\% & 90.72\% & 90.74\% & - \\
		 \multicolumn{2}{|c|}{} & 1 & 100  & 93.37\% & - & - & - \\
		 \multicolumn{2}{|c|}{} & 3 & 50  & 97.16\% & 97.18\% & 97.04\% & - \\
		\hline
		\multicolumn{2}{|c|}{Ethernet (Downlink)} & 3 & 100  & 92.47\% & 91.64\% & 90.79\% & - \\
		\hline
		\multicolumn{2}{|c|}{Cached} & 3 & 100  & 95.88\% & 95.30\% & 95\% & - \\
		\hline
		\multirow{2}{*}{Slotted \quad} & 1ms & 3 & 100  & 89.23\% & 88.25\% & 87.98\% & - \\
		 & 10ms & 3 & 100  & 63.73\% & 61.40\% & 63.35\% & - \\
		\hline
		\multicolumn{2}{|c|}{Femtocell} & 3 & 100 & 92.60\% & 91.80\% & 91.83\% & - \\
		\hline
		\multicolumn{2}{|c|}{\multirow{1}{*}{Tor}} & 3 & 100  & 58.44\% & 56.18\% & 56.2\% & -  \\
		\hline

	\end{tabular}
	\caption{Summary of the measured success rate of the proposed attack reported here.  Data is shown for different numbers of exemplars, different population sizes and different values of K in the K-nearest neighbours method. In all cases the samples of each web site are fetched consecutively within an hour except for $(^*)$ where a sample is taken each hour for 5 days.}
	\label{tab:knn}
\end{table*}

\subsection{Other Proposed Channels}
\noindent
A number of other channels have been proposed in the literature as a defence against traffic analysis attacks.  Wright \emph{et al} \cite{wright09} suggest a traffic morphing method which maps the packet sizes of one web site to the packet distribution of another site. This defence fails to overcome the attack considered here since it makes use only of timing information and does not use packet size information. This is also the case for all of the packet-size based defences proposed in the HTTPOS scheme introduced in  \cite{luo11}.  A potential defence against timing attacks is to modify the packet timing pattern by delaying transmissions.  However, although this might be expected to counter timing-based attacks such as that considered here such defences will also have an impact on delay.  For example, BuFLO introduced in \cite{dyer12} is similar to the time slotting method that we consider above and which appears to be impractical given its substantial impact on delay and bandwidth, with $190\%$ bandwidth overhead reported in \cite{tao14}.

\section{Effect of Link Speed and Content Change on Classification Performance}
\label{sec:content_change}
\noindent
{By looking closely at performance of websites, it can be seen that the total mean success rate obtained in each measurement campaign is not monotone amongst individual websites. In this section, we investigate possible reasons behind the poor performance of certain websites.	We use the same ethernet dataset from Section \ref{sec:ether} where samples are fetched hourly over 5 days. The study of other scenarios like femtocell, cached \etc. provides similar results.}

\begin{figure}[!t]
\centering
\subfloat[Open IP connections]{
  \includegraphics[trim=1cm 0cm 1cm 0cm,clip, width=0.5\columnwidth]{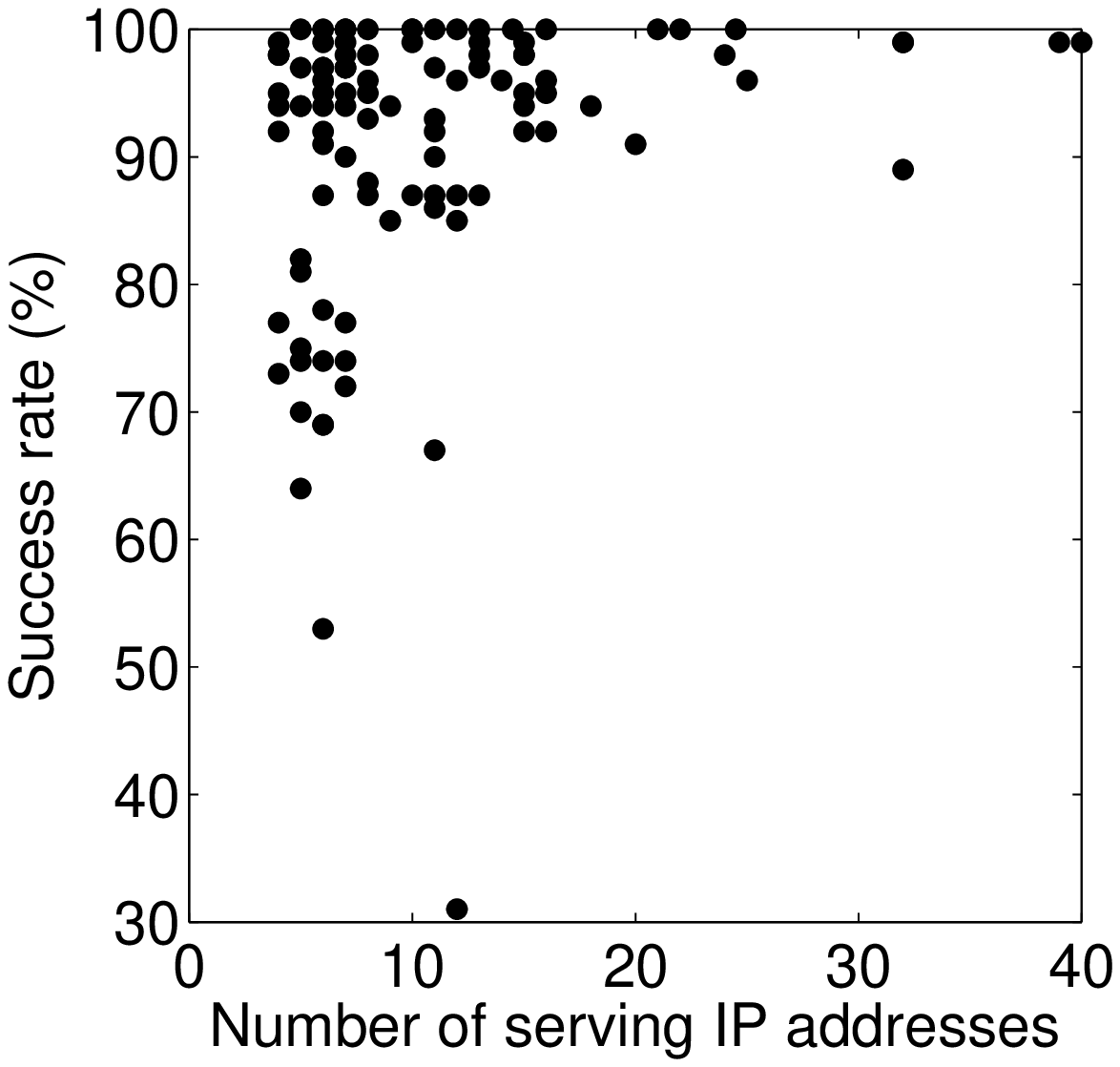}
  \label{fig:medips}
}
\subfloat[Active TCP ports]{
  \includegraphics[trim=1cm 0cm 1cm 0cm,clip, width=0.5\columnwidth]{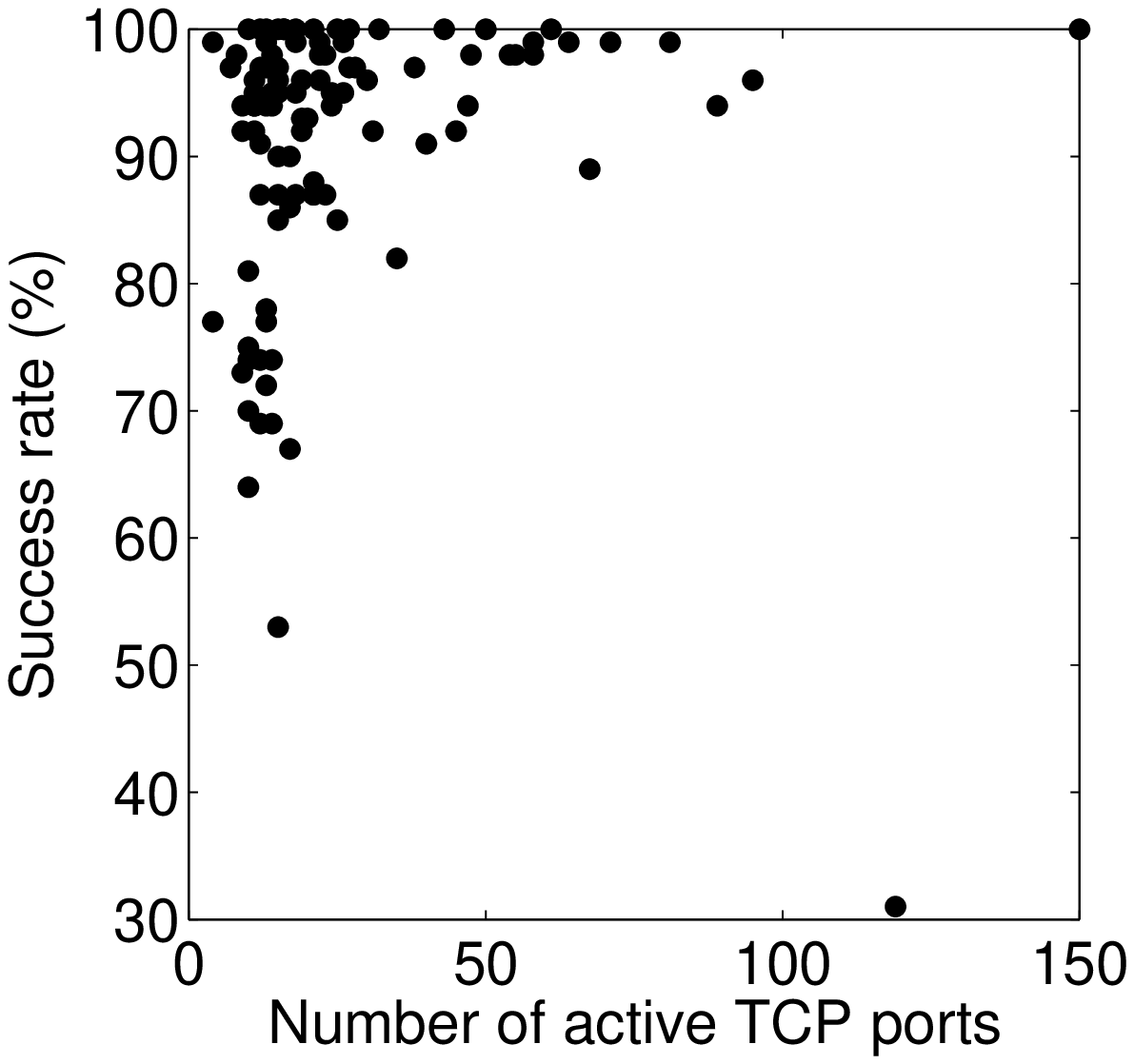}
  \label{fig:medports}
}
\caption{{Scatter plot of median open IP connections and active ports count against success rate. Samples are taken hourly for 5 days over ethernet channel.}}
\end{figure}

\begin{enumerate}
\begin{figure}[!htb]
\centering
\subfloat[Standard deviation]{
  \includegraphics[trim=1cm 0cm 1cm 0cm,clip, width=0.5\columnwidth]{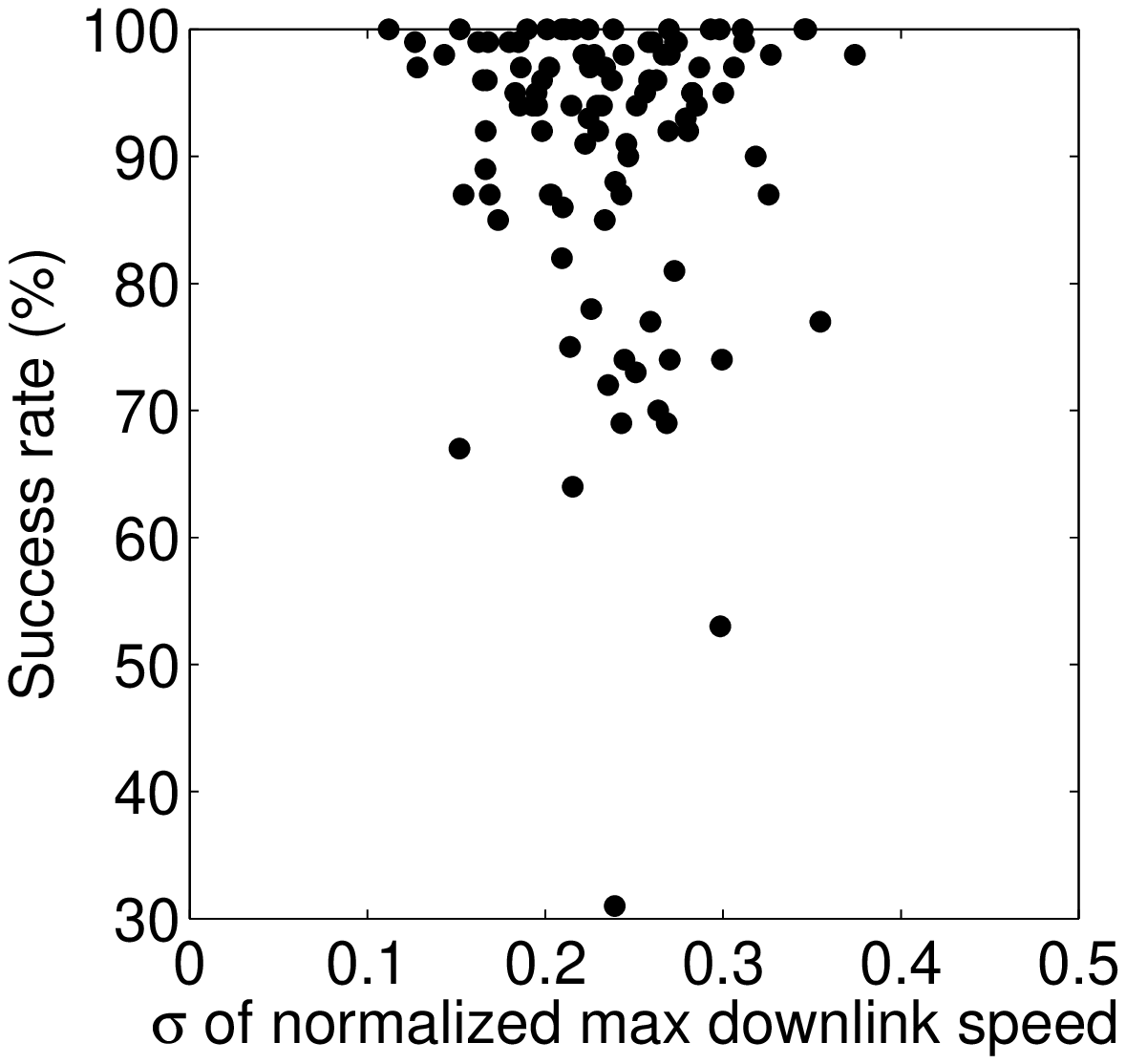}
  \label{fig:stdspeed}
}
\subfloat[Median]{
  \includegraphics[trim=1cm 0cm 1cm 0cm,clip, width=0.5\columnwidth]{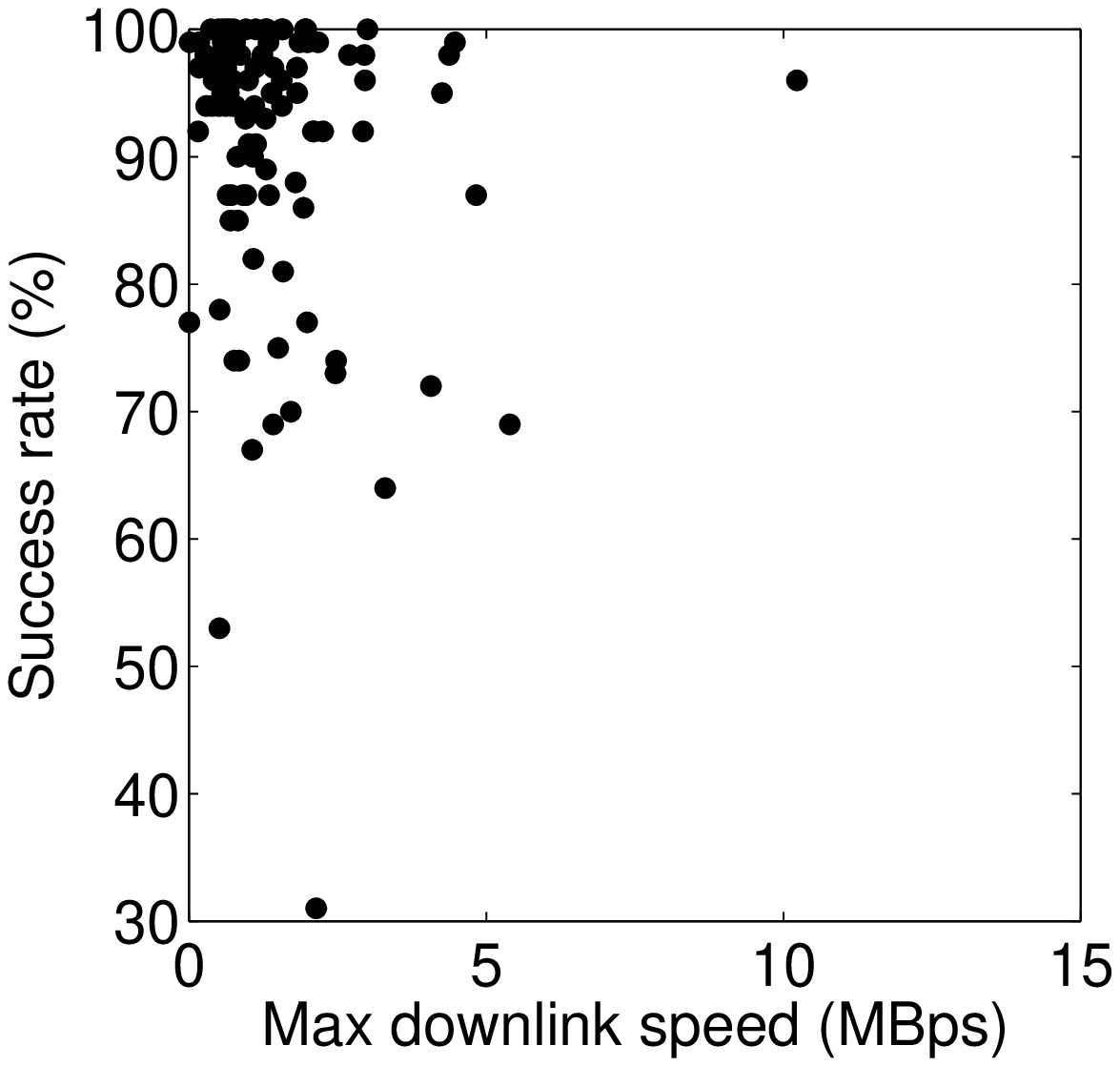}
  \label{fig:medspeed}
}
\caption{{Scatter plot of max link speed standard deviation and median against success rate. Samples are taken hourly for 5 days over ethernet channel.}}
\end{figure}

\item {\textit{Network Speed.} The link speed between the client and each web server varies from a website to another. It is also different between samples of the same page. To investigate the effect of network speed on the classification performance, we calculated peak downlink speed during each fetch (the results for uplink and uplink+downlink speed is similar). Then in order to compare the metrics, values for samples of each page are normalized and their variance is evaluated. Figure \ref{fig:stdspeed} illustrates the scatter plot of normalized standard deviation of link speed against success rate of each website. It can be seen there is no strong correlation between these two metrics that would suggest that a web site with more variable link speed should result a lower success rate. Similar comparison is also studied with median speed for each web site (Figure \ref{fig:medspeed}) to show that having an overall faster link speed does not guarantee a poor classification performance.}

\begin{figure}[!htb]
\centering
\subfloat[Sample length]{
  \includegraphics[trim=1cm 0cm 1cm 0cm,clip, width=0.5\columnwidth]{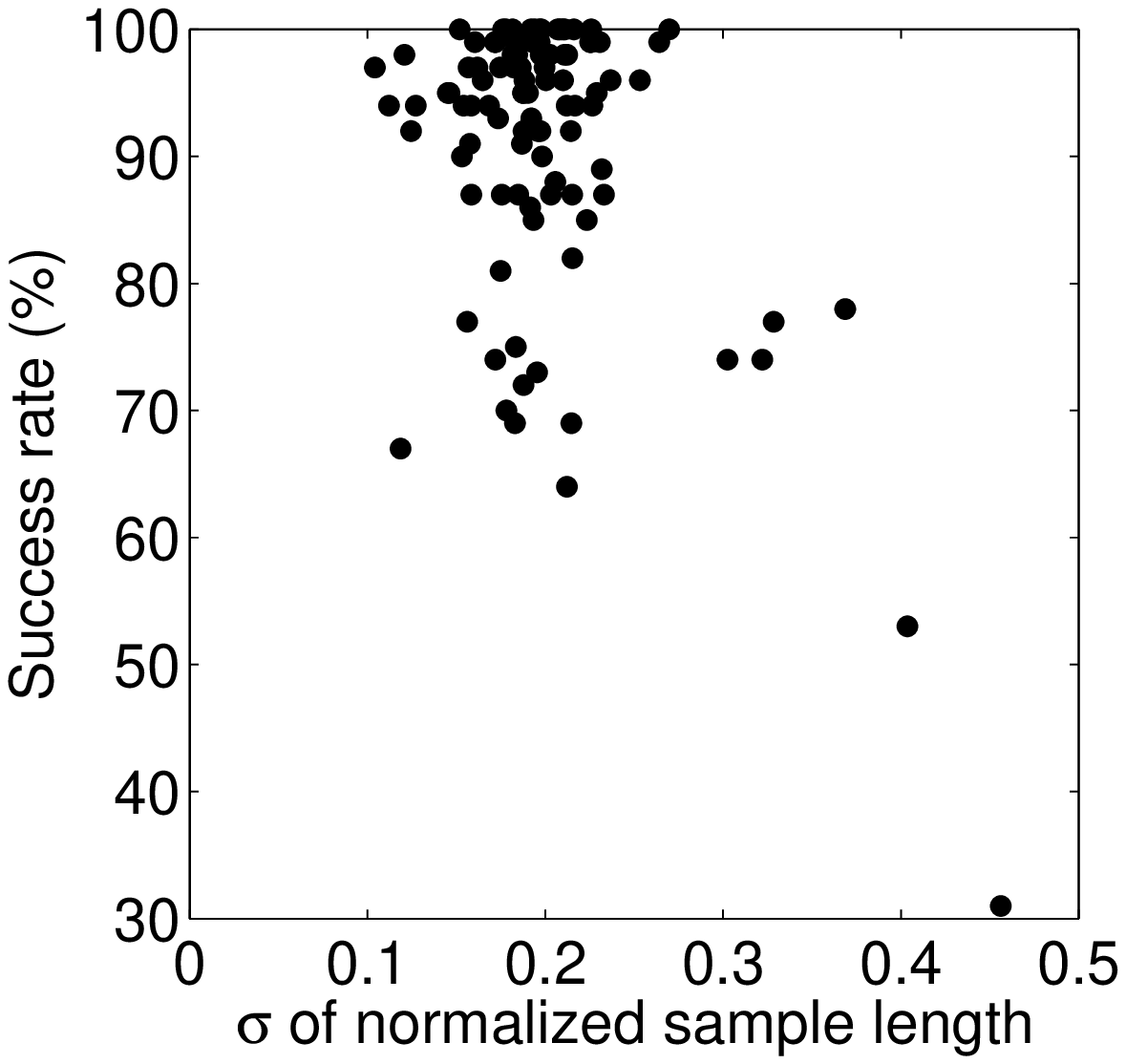}
  \label{fig:stdlength}
}
\subfloat[GET/POST count]{
  \includegraphics[trim=1cm 0cm 1cm 0cm,clip, width=0.5\columnwidth]{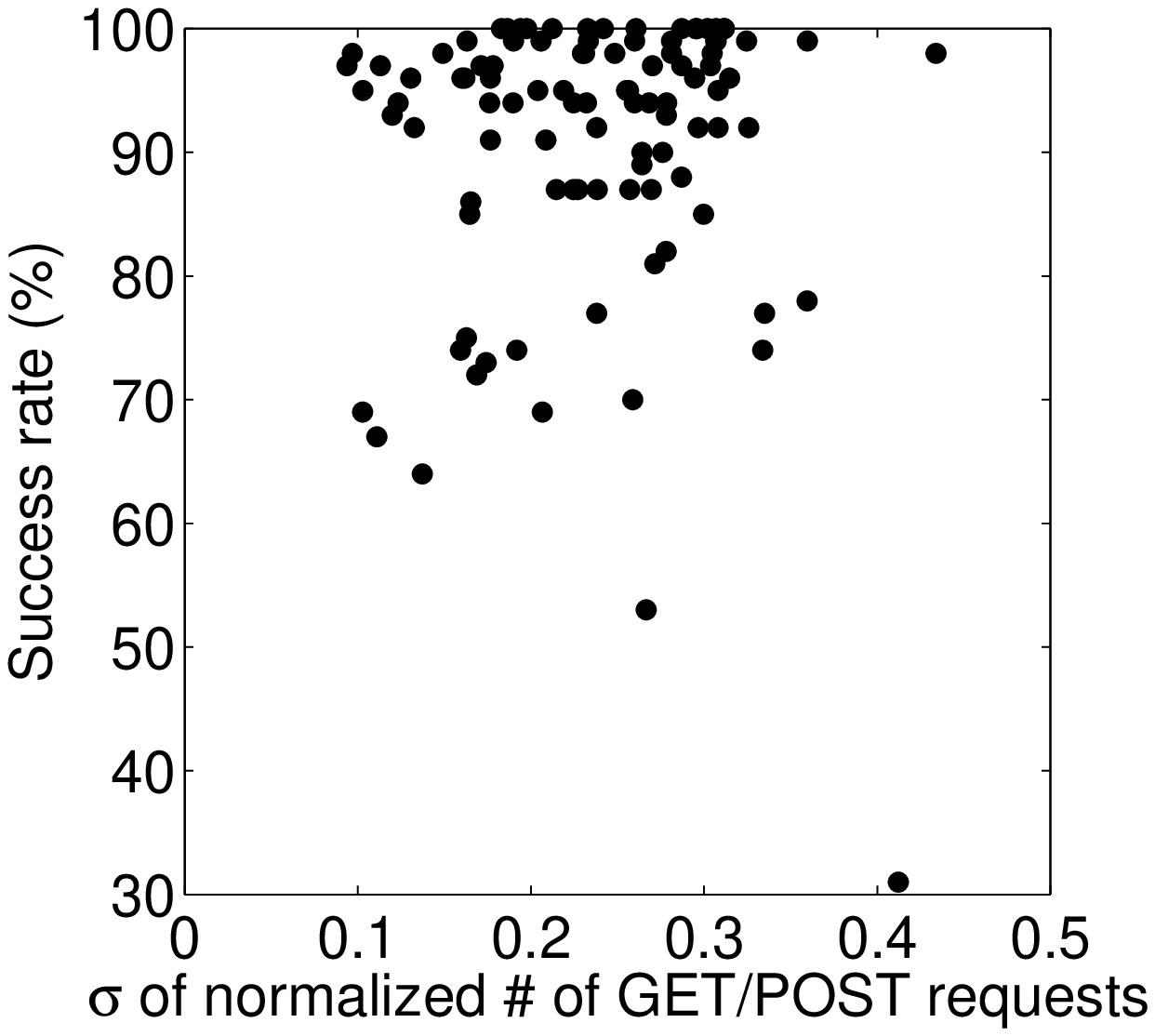}
  \label{fig:stdgets}
}
\caption{{Scatter plot of sample length and GET/POST request count standard deviation against success rate. Samples are taken hourly for 5 days over ethernet channel.}}
\end{figure}

\item {\textit{Sample Length and GET/POST Requests Count.} For each web site we plot the standard deviation of the normalized number of uplink packets (a measure of the variability of the web page over time) and the corresponding success rate (see Figure \ref{fig:stdlength}). The results for uplink and uplink+downlink is similar. We also provided the same plot for maximum number of GET/POST requests for each website (Figure \ref{fig:stdgets}). It can be seen that, there is no strong correlation between the these metrics and success rates which is suggestive that the classification attack is fairly insensitive to variability of web page content over time.}

\item {\textit{IP Connections, Active TCP ports.} In order to investigate the robustness of the attack against parallel connections, for each web site we plot the median number of serving IP connections and active TCP ports against their corresponding success rates. As illustrated in Figures \ref{fig:medips} and \ref{fig:medports}, again there is no clear correlation between mentioned metrics which is suggestive that the number of active IPs/ports for each web site, which represents the number of parallel connections, has no effect on the performance of our proposed attack.}
\end{enumerate}

The above results suggest that there is no strong correlation between the performance of our attack and link speed, small content change and number of parallel connections. However the choice of exemplars are essential to the performance of the attack.	In particular when the content change is more than a threshold, the difference between samples can no longer be ignored by the attack.	An example of this misbehaviour can be seen for website $\#10$ in the measurement campaign considered in this section, where 2 different versions of the page were observed during the experiment. In result, 1 exemplar represents one version while 2 others represent another version of the page. This causes $K$-NN method to fail collecting enough votes for a successful classification, which in turn leads to a success rate of $31\%$.

To overcome this issue, separate sets of exemplars are required to represent each version of a web page in order to successfully classify future samples.

\section{Finding a Web Page within a Sequence of Web Requests}
\label{sec_seq}
\noindent
In the experiments presented so far we have assumed that within the observed packet timestamp stream the boundaries between different web fetches are known.   This is probably a reasonable assumption on lightly loaded links where the link is frequently idle between web fetches.	However, not only  might this assumption be less appropriate on more heavily loaded links but it also allows for a relatively straightforward means of defence, namely insertion of dummy packets to obscure the boundaries between web fetches.   In this section we therefore extend consideration to links where web fetches are carried out in a back to back fashion such that the boundaries between web fetches cannot be easily identified.   

The basic idea is to sweep through a measured stream of packet timestamps trying to match sections of it against the timing signature of a web page of interest.  This exploits the fact that our timing-only attack does not fundamentally depend on knowledge of the start/end times of the web fetch (unlike previous approaches which use packet counts to classify web pages).   

In more detail, to locate a target web page within a stream of packet timestamps we first select three measured packet timestamp sequences for that web page to act as exemplars (as previously). 
Then, we sweep through the stream of timestamps in steps of 10 packets, extract a section of the stream of the same length as each exemplar (plus 10 to cover the step size) and calculate the distance between the section and the exemplar.  After sweeping through the full stream we select the location within the stream with least distance from the exemplars as the likely location of the target web page within the stream.   While this process assumes that the target web page is present within the packet stream, using a similar approach to that in Section \ref{sec:falsepos} we could extend this approach to decide whether the web page is present by appropriately thresholding the distance (when the measured least distance is above the threshold, the page is judged to not be present in the stream).    

\subsection{Results}
\begin{figure*}[!t]
\centering
\subfloat[Two consecutive web fetches]{
  \includegraphics[trim=0cm 1cm 0cm 2cm,clip, width=0.9\columnwidth]{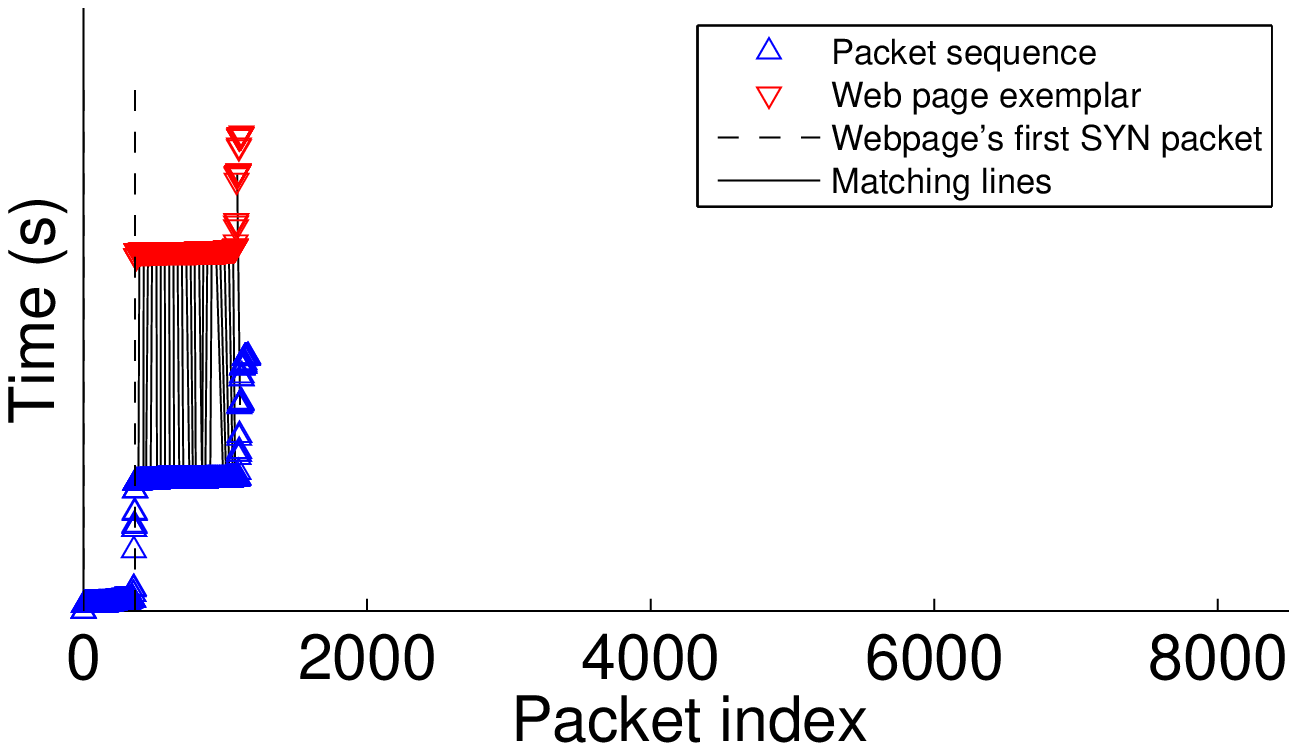}
}
\subfloat[Three consecutive web fetches]{
  \includegraphics[trim=0cm 1cm 0cm 2cm,clip, width=0.9\columnwidth]{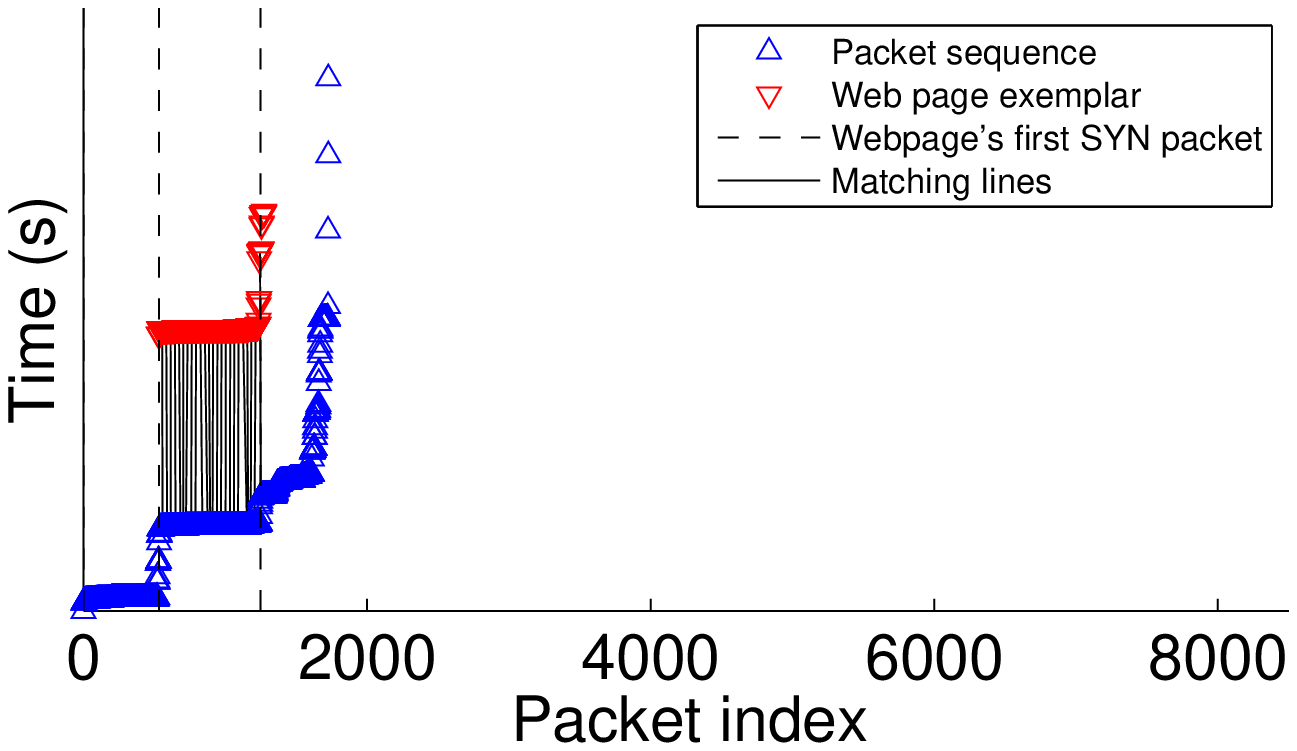}
}
\hspace{0mm}
\subfloat[Four consecutive web fetches]{
  \includegraphics[trim=0cm 1cm 0cm 2cm,clip, width=0.9\columnwidth]{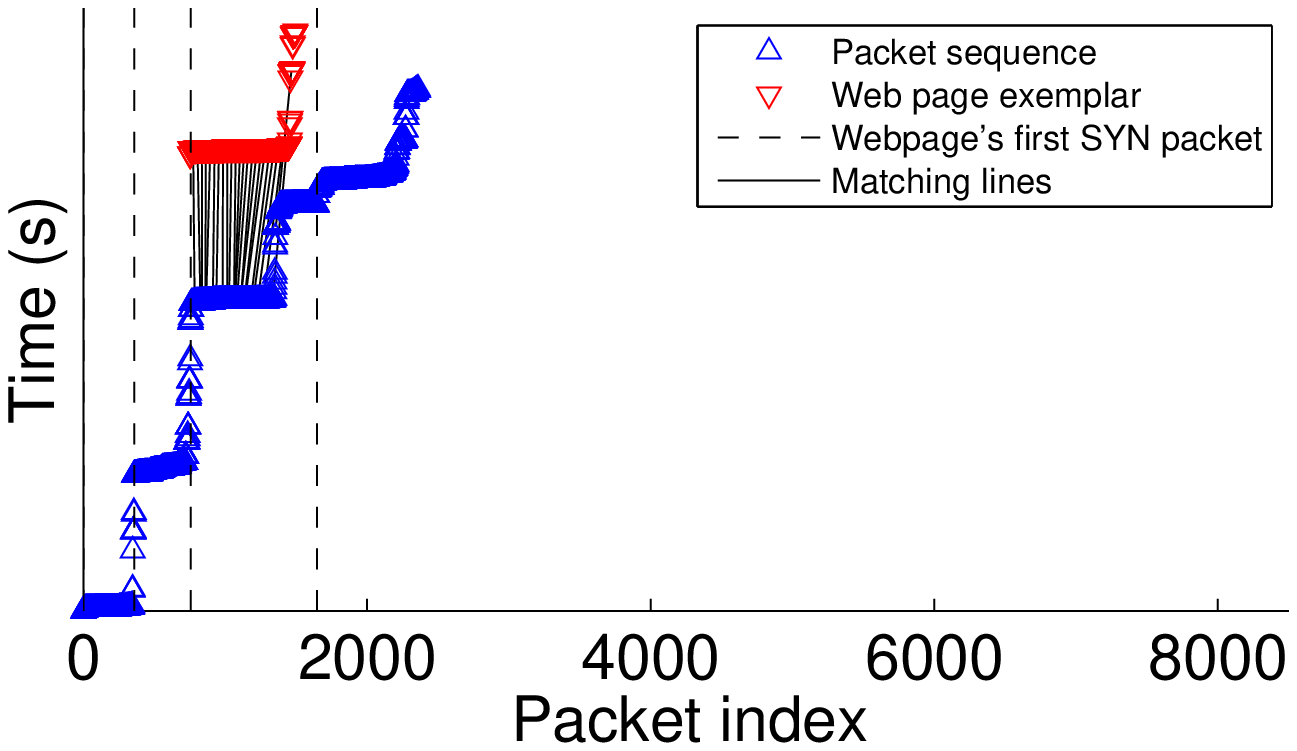}
}
\subfloat[Five consecutive web fetches]{
  \includegraphics[trim=0cm 1cm 0cm 2cm,clip, width=0.9\columnwidth]{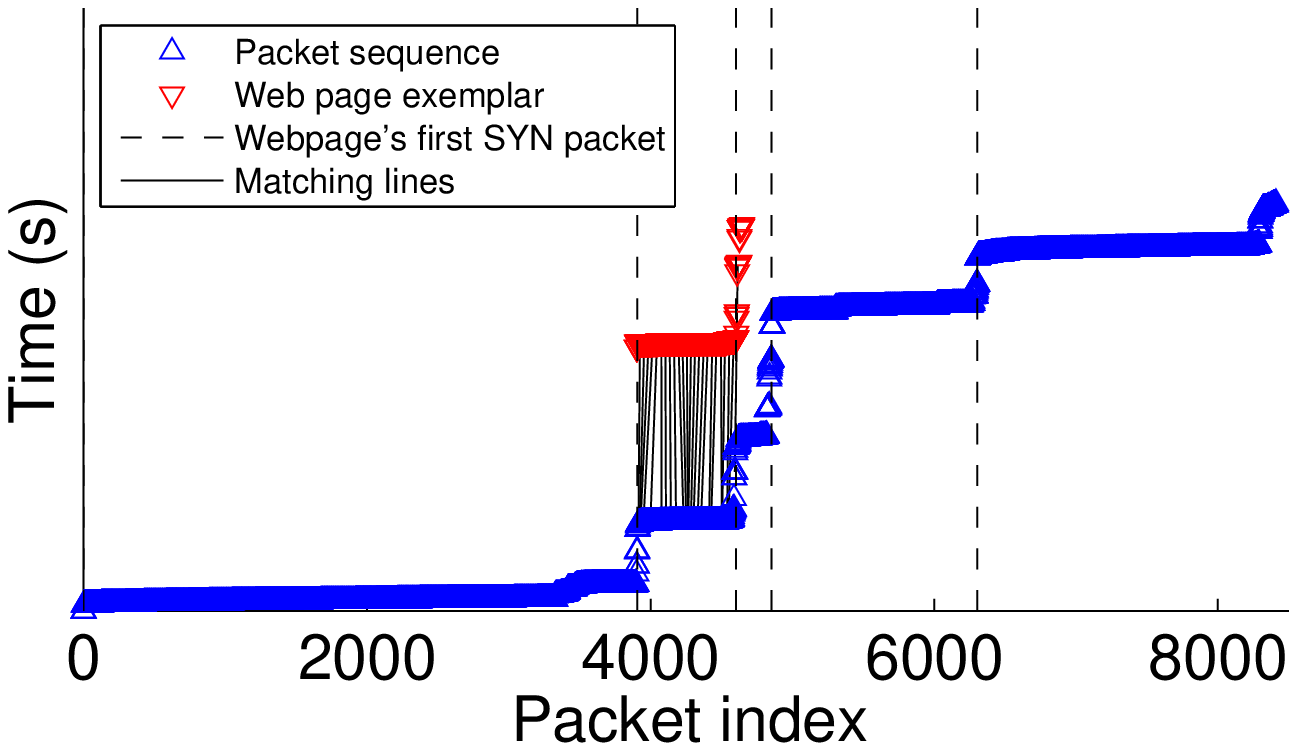}
}
\caption{Illustrating locating of a web page within a packet stream. The page www.iscp.ie shown in red triangles is an example of a web page which is successfully located among 2, 3, 4 and 5 consecutive web fetches. The vertical lines show the first SYN packet of each web page.}
\label{fig_multi}
\end{figure*}
\noindent
We constructed a test dataset as follows.  For each run we pick one of the 100 web sites to be the target. We then uniformly at random pick up to 4 other web sites from the remaining web sites. The selected web sites are then permuted randomly and fetched one after another with a pause after each fetch acting as a ``thinking period''. The maximum time allowed for each fetch to complete is $25$ seconds \emph{i.e} the length of each pause is selected uniformly at random from \mbox{$5$-$25$} seconds.  Repeating this for all web sites in the dataset, we created 100 test runs. 

\begin{table}[!htb]
\centering
	\begin{tabular}{| l | c | c | c | c |}
		\hline
		No. of consecutive pages & 2 & 3 & 4 & 5 \\
		\hline
		Success rate & 82\% & 80\% & 66\% & 64\% \\
		\hline
	\end{tabular}
	\caption{Success rates of locating web pages among \mbox{$2$-$5$} fetches.}
	\label{tab:stream}
\end{table}

Using the classification approach described above we attempted to identify the location within each packet stream.	Figure \ref{fig_multi} presents four examples of this, showing the position within a stream with least distance from the exemplars of a target web page.	The success rate results for streams of \mbox{$2$-$5$} web sites are summarized in Table \ref{tab:stream}.	With this approach we achieved a maximum success rate of $82 \%$ for  locating the target web page within each packet stream within a position error of $w . l_s$ packets, where $w$ is the window size at which DTW operates ($0.2$ in our setting) and $l_s$ is the average length of the $3$ exemplars which are determined for each web site $s$ separately.   Given the limited information being used, this is a remarkably high success rate and indicates the power of the timing-only attack.   However, it can be seen that the success rate starts to lower as the number of consecutive fetches grows which leads to a longer packet stream that can potentially include similar patterns to the target web page. Moreover web pages with shorter length are less likely to be located properly due to their shorter signatures which are more likely to appear in the middle of a larger web trace.

\section{Summary and Conclusions}
\noindent
We introduce an attack against encrypted web traffic that makes use only of packet timing information on the uplink.   In addition, unlike existing approaches this timing-only attack does not require knowledge of the start/end of web fetches and so is effective against traffic streams.   We demonstrate the effectiveness of the attack against both wired and wireless traffic, consistently achieving mean success rates in excess of 90\%.    Table \ref{tab:knn} summarises our measurements of the success rate of the attack over a range of network conditions.

Study of downlink and a preliminary study of uplink+downlink traffic suggest little difference from uplink results presented in this paper, given timing patterns of uplink and downlink are strongly correlated. Moreover, the proposed attack proves to be robust against different link speed, different number of parallel connections and small content change, being able to maintain overall success rate of $91\%$ for measurements collected over a course of 5 days. However the threshold for which the attack remains resilient to content change is to be studied. we leave further investigation of these matters for future work.

Since this attack only makes use of packet timing information it is impervious to existing packet padding defences.   We show that time slotting is also insufficient to prevent the attack from achieving a high success rate, even when relatively large time slots are used (which might be expected to significantly distort packet timing information).   Similarly, randomised routing as used in Tor is also not effective.   More sophisticated types of defence may be more effective, but we leave consideration of those to future work as they likely involve complex trade-offs between network performance (\emph{e.g.} increased delay and/or reduced bandwidth) and resistance to attack that warrant more detailed study than is possible here.

In addition to being of interest in its own right, by highlighting deficiencies in existing defences this timing-only attack points to areas where it would be beneficial for VPN designers to focus further attention.

\balance
\bibliography{references}{}
\bibliographystyle{plain}

\begin{IEEEbiography}{Saman Feghhi}

is pursuing a PhD degree in Computer Science at School of Computer Science and Statistics in Trinity College Dublin, Ireland. He received his master's and bachelor's degrees also in Computer Science from Sharif University of Technology in Iran. His current research interests are computer networks, internet privacy, network security and mobile network data analytics.

\end{IEEEbiography}

\begin{IEEEbiography}{Douglas J. Leith}

graduated from the University of Glasgow in 1986 and was awarded his PhD, also from the University of Glasgow, in 1989. In 2001, Prof. Leith moved to the National University of Ireland, Maynooth and then in Dec 2014 to Trinity College Dublin to take up the Chair of Computer Systems in the School of Computer Science and Statistics.  His current research interests include wireless networks, network congestion control, distributed optimisation and data privacy.   

\end{IEEEbiography}

\end{document}